\documentclass{aa}
\usepackage{epsf,times,latexsym}

\usepackage{graphicx,epsfig}
\newcommand{\bq}{\begin{equation}}
\newcommand{\eq}{\end{equation}}
\newcommand{\bqn}{\begin{eqnarray}}
\newcommand{\eqn}{\end{eqnarray}}

\titlerunning{Simulations of unstable modes}

\authorrunning{A.V. Khoperskov et al.}

\title
{High resolution simulations of unstable modes in a collisionless disc }
\author{A. V. Khoperskov \inst{1} \and A. Just\inst{2}
   \and V. I. Korchagin \inst{3} \and M. A. Jalali\inst{4}
}

\offprints{A. Just} \mail{just@ari.uni-heidelberg.de} \institute{
Department of Theoretical Physics, Volgograd State University,
Volgograd, 400068, Russia; Isaac Newton Institute of Chile, Moscow
Branch \and Astronomisches Rechen-Institut at ZAH, University of
Heidelberg, M\"onchhofstra{\ss}e 12-14, 69120 Heidelberg, Germany
\and Institute of Physics, Stachki 194, Rostov-on-Don, 344090,
Russia; Isaac Newton Institute of Chile, Rostov-on-Don Branch \and
Department of Mechanical Engineering, Sharif University of
Technology, Azadi Ave., Tehran, Iran }

\begin{document}

\date{Printed: \today}

\maketitle

\abstract
{We present $N$-body simulations of unstable spiral modes in a dynamically
cool collisionless disc. We show that spiral modes grow in a thin
collisionless disk in accordance with the analytical perturbation theory.
We use the particle-mesh code {\sc Superbox} with nested grids to follow
the evolution of unstable spirals that emerge from an unstable equilibrium
state. We use a large number of particles (up to $N=40\times 10^6$) and
high-resolution spatial grids in our simulations ($128^3$ cells).
These allow us to trace the dynamics of the unstable spiral modes until
their wave amplitudes are saturated due to nonlinear effects.
In general, the results of our simulations are in agreement with the analytical
predictions. The growth rate and the pattern speed of the most unstable
bar-mode measured in $N$-body simulations agree with the linear analysis.
However the parameters of secondary unstable modes are in
lesser agreement because of the still limited resolution of our
simulations.

 }

\section{Introduction\label{sec-intro}}

Gravitational instability of galactic discs is a widely accepted
physical mechanism for the generation of barred and spiral structures
in disc galaxies. Since the pioneering publications by Lin \& Shu (1964)
and Toomre (1981), considerable progress has been made in the theoretical
study of gravitational instabilities of disc-like systems. A steady stream
of papers performing a linear stability analysis of self-gravitating
gaseous discs (e.g., Bertin et al. 1989a,b; Adams et al. 1989; Savonije \&
Heemskerk 1990) have demonstrated that a massive gaseous disc will almost
inevitably be prone to spiral instabilities. A number of numerical
techniques have been applied to study the nonlinear evolution
of unstable gaseous discs (e.g., Tomley et al. 1994; Miyama et al. 1994;
Woodward et al. 1994; Nelson et al. 1998; Laughlin et al. 1997,1998).
Recently, the linear approach has been generalized for the stability
analysis of non-axisymmetric gaseous disks (Asghari \& Jalali 2006).
These contributions address the dynamical modeling of gaseous
discs with a variety of equilibrium properties and compare the results
of numerical simulations with those of analytical predictions.

The dynamical study of the collisionless gravitating discs is
hampered by the difficulty of normal mode calculations in such
systems. A global modal analysis of collisionless discs has been
accomplished for a few cases. Kalnajs (1972) calculated the global
modes of a uniformly rotating, self-gravitating stellar disc. Zang
(1976) carried out a modal analysis of Mestel's stellar disc using
an inner cutout function to handle the central singularity. His
method was then used by Evans \& Read (1998a,b) for a general class
of scale-free discs with rising and falling rotation curves. The
modal properties of a collisonless exponential disc (with a core)
were also studied by Vauterin \& Dejonghe (1996). Most recently,
Jalali \& Hunter (2005, hereafter JH) developed a rather general
approach that allowed them to calculate the modal properties of
stellar discs. They showed that the radial orbits of soft-centered
models lead to a boundary integral that plays a crucial role in the
formation and growth of a bar mode, which is localized within the
central regions of a collisionless disc. They also investigated
the effect of dark halos/bulges on the properties of growing modes.

The major drawback of perturbation theories is that they cannot
make any prediction of modal evolution in the nonlinear regime.
A linear analysis cannot predict the saturation level of the wave
amplitude after an exponential growth phase, nor can it predict
the duration of the linear and nonlinear phases. One can tackle
these problems by using $N$-body simulations.

An $N$-body approach has a number of difficulties, such as numerical
relaxation due to a high level of noise and finite spatial resolution.
These factors hamper the behavior of collisionless models. Moreover,
growing modes are sensitive to the equilibrium profiles of the central
regions. Thus, numerical simulations with a large number of particles,
and high spatial resolution, are needed to correctly model the behavior
of unstable modes in collisionless discs.

There have been a few attempts to numerically model the dynamics of
global modes in collisionless discs. As already mentioned,
modeling of the spiral structure in collisionless galaxies
and comparison with analytical predictions has been restrained
by the relatively small number of particles used in $N$-body codes.
Randomly generated initial coordinates of particles inevitably lead to
density fluctuations and perturbing forces that affect the
disc dynamics (Sellwood 1983). The disc response to these virtual
forces could contribute substantial errors of about 50 percent to
the computed growth rates and the pattern speeds (Sellwood 1983).
In previous $N$-body simulations aimed at studying the dynamics
of global modes in collisionless discs
(Athanassoula \& Sellwood 1986; Sellwood \& Athanassoula 1986),
special precautions have been taken to avoid a strong influence
of noise perturbations. These authors used `quiet'
initial conditions, with the groups of stars spaced at
equal intervals around the rings at fixed radii, and all particles
in a ring were given the same initial radial and tangential
velocity components.

The development of numerical techniques with tens of millions of
particles would avoid artificial precautions in tackling
the numerical noise problem. In this paper, we use {\sc Superbox}
(Fellhauer et al. \cite{fel00}) to model the evolution of a
collisionless disc. {\sc Superbox} is a highly efficient
particle-mesh-code with nested and co-moving grids, based on
a leap-frog scheme designed for the simulation of interacting
galaxies or other stellar systems. Aiming at modeling the
dynamics of spiral patterns in real galaxies, we apply the code
to analyse one of JH models that has an exponential disc density
distribution with a core, embedded in a cored logarithmic potential.
We use this model for two reasons. Firstly, closed-form expressions
are available for the phase space distribution functions. Secondly,
a full spectrum of unstable modes has been computed for this model.

We find good agreement between the linear global modal analysis
and the nonlinear simulations. The growth rate and the pattern speed
of the most unstable $m$=2 bar-like global mode found in $N$-body
simulations, and the spatial distribution of this mode, agree well with
the analytical results. We have also been able to recover the
theoretical growth rates and the pattern speeds of the unstable
$m$=3 and $m$=4 global modes. Apart from demonstrating the existence
of global modes in the cored exponential
discs, our simulations provide a welcome check of the {\sc Superbox}
code, and demonstrate its applicability to model the dynamics of
real galaxies.

In \S\ref{sec-ana} we briefly summarize the properties of the global
modes of JH models, and use their method to find the unstable $m$=2,
$m$=3 and $m$=4 modes of a cored exponential disc. In
\S\ref{sec-num} we describe the $N$-body code and its results for
the model of JH. A comparison of the results of $N$-body simulations
with those of analytical predictions is given in the same section.
\S\ref{sec-disc} contains a summary of our results.
\begin{figure}
\centerline{
  \resizebox{0.98\hsize}{!}{\includegraphics[angle=270]{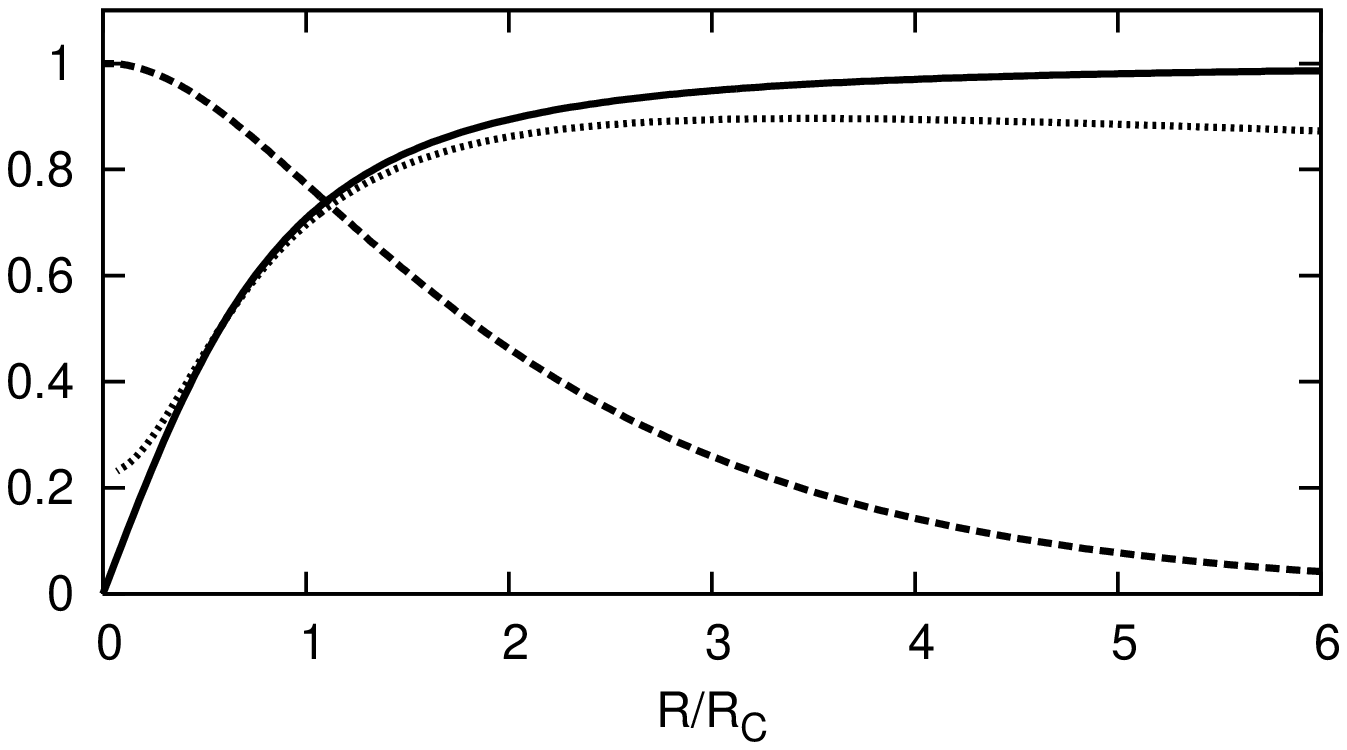}}
  }
\centerline{
  \resizebox{0.98\hsize}{!}{\includegraphics[angle=270]{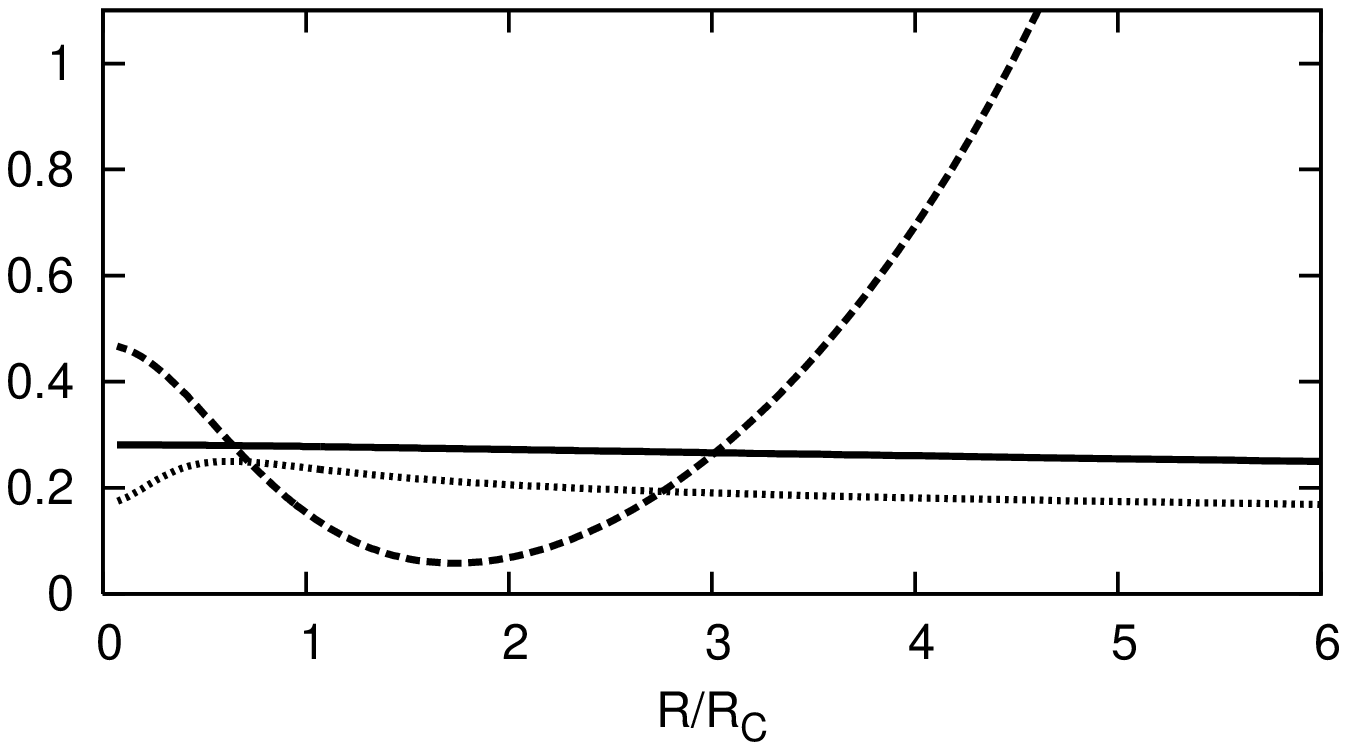}}
  }
\caption[]{{\it Upper panel}: Solid line shows the rotation curve
$v_{\rm rot}$ of the model. Dotted line is the streaming velocity
$\langle v_\varphi \rangle$ and dashed line is the normalized surface
density distribution. {\it Lower panel}: Solid and dotted lines show
$C_R$ and $C_\varphi$, respectively. Dashed line is Toomre's $Q$ minus
1 ($Q-1$). We have set the model parameters to $N$=6, $G\Sigma_0
R_D/v_0^2$=0.34 and $\lambda=R_C/R_D$=0.625.} \label{fig:fig1}
\end{figure}
\begin{figure*}
\centerline{\hbox{\epsfxsize=8.5truecm\epsfbox{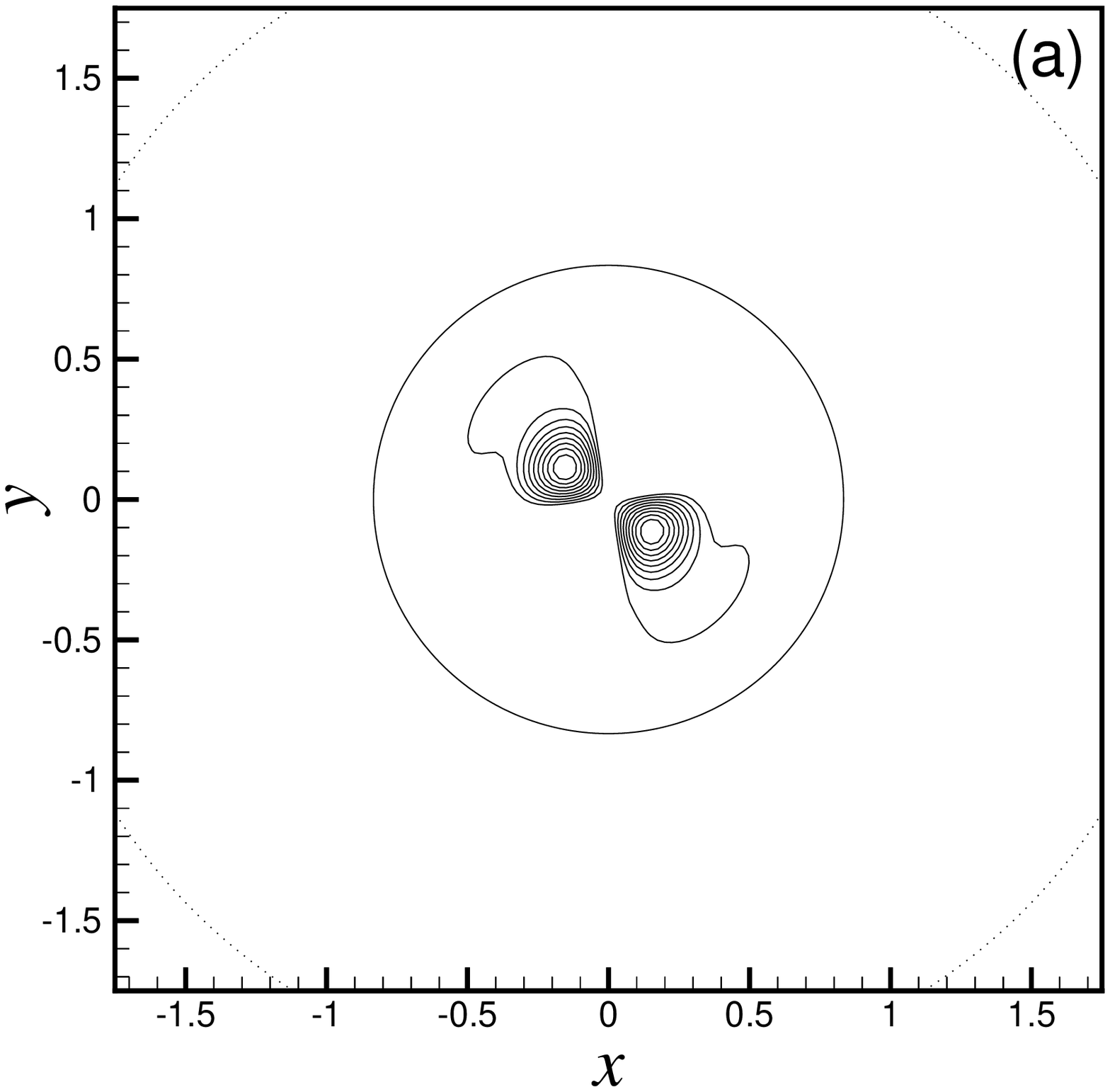}
            \hspace{0.2truecm}
                  \epsfxsize=8.5truecm\epsfbox{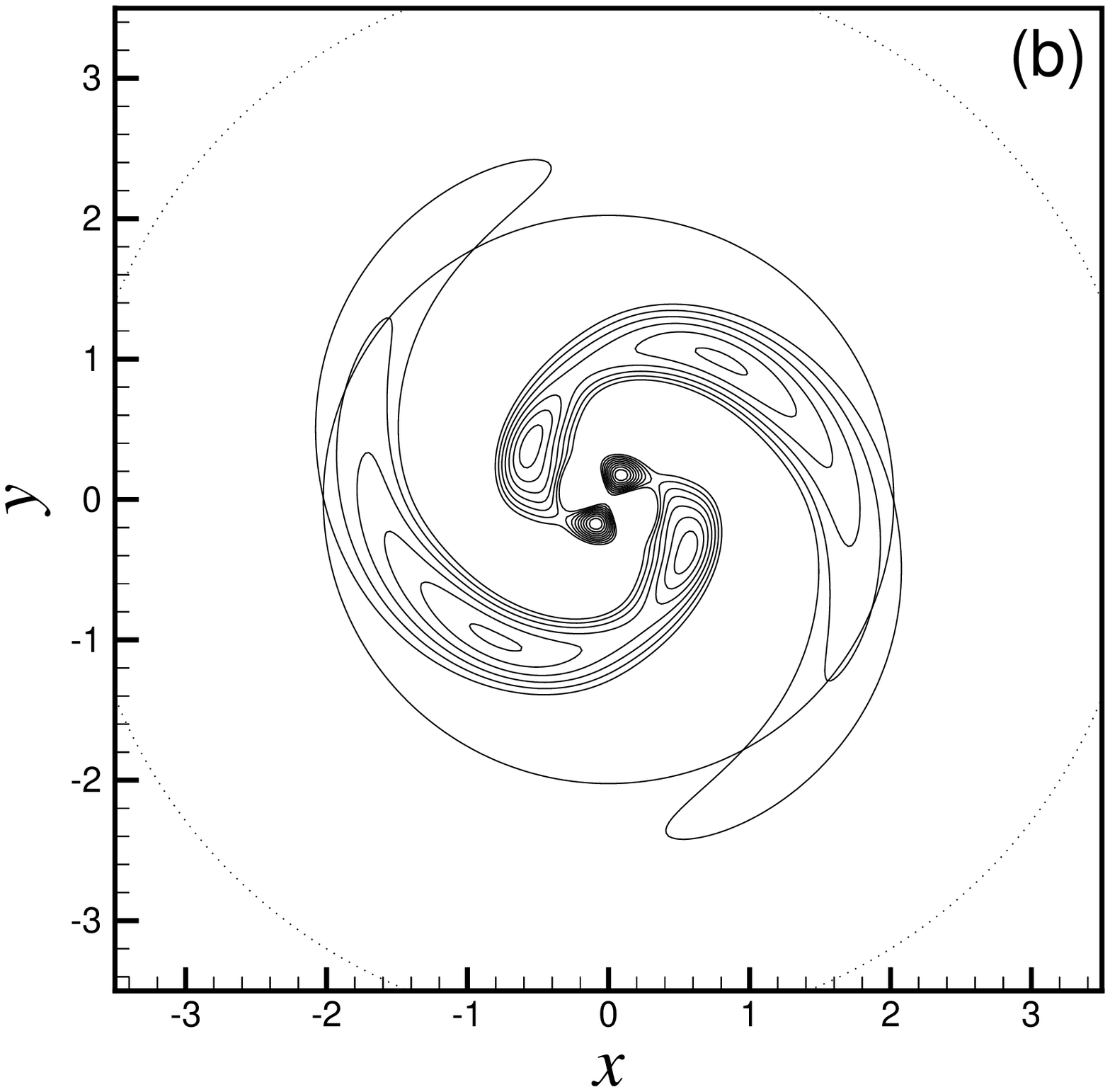}}}
\centerline{\hbox{\epsfxsize=8.5truecm\epsfbox{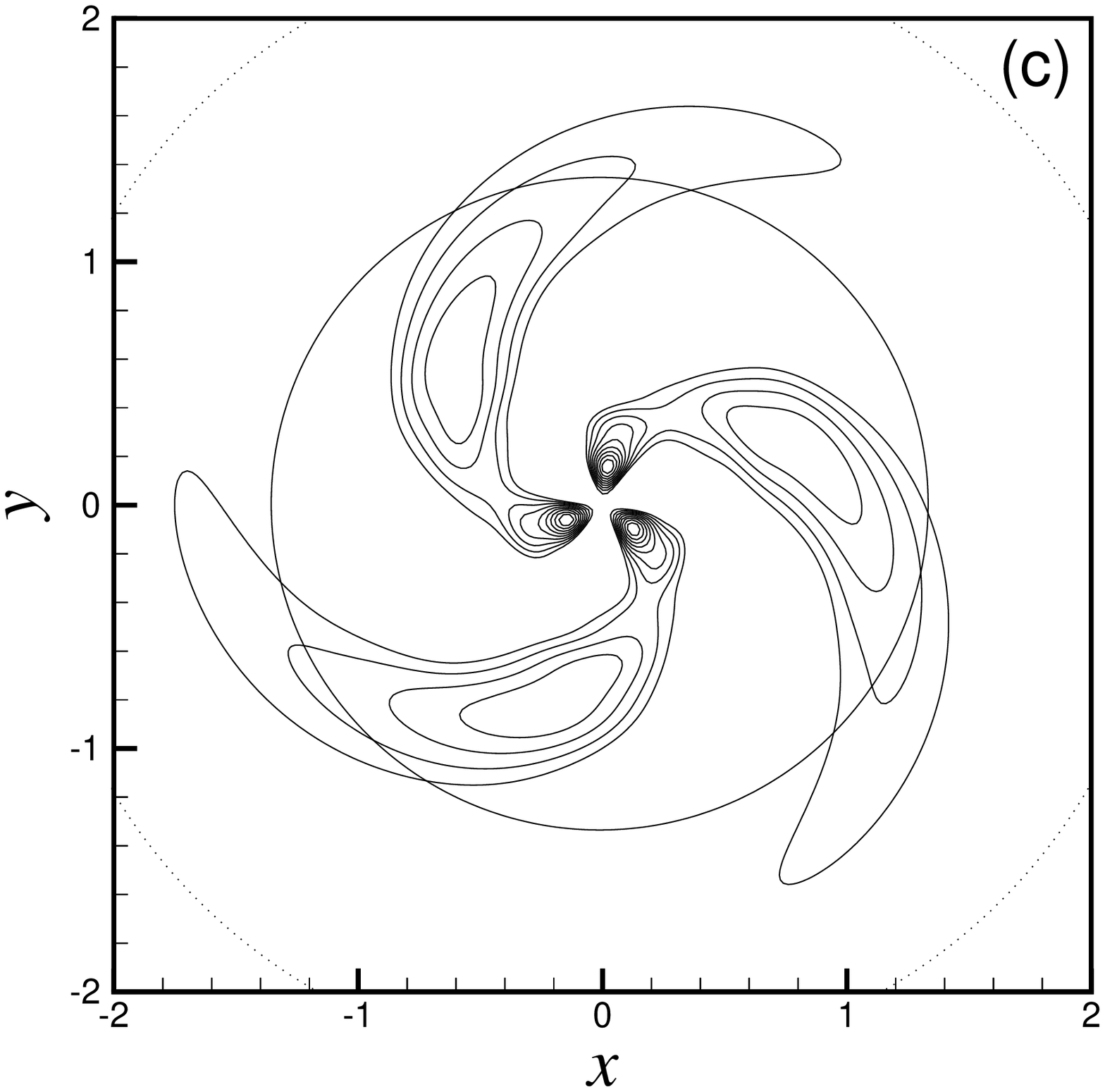}
            \hspace{0.2truecm}
                  \epsfxsize=8.5truecm\epsfbox{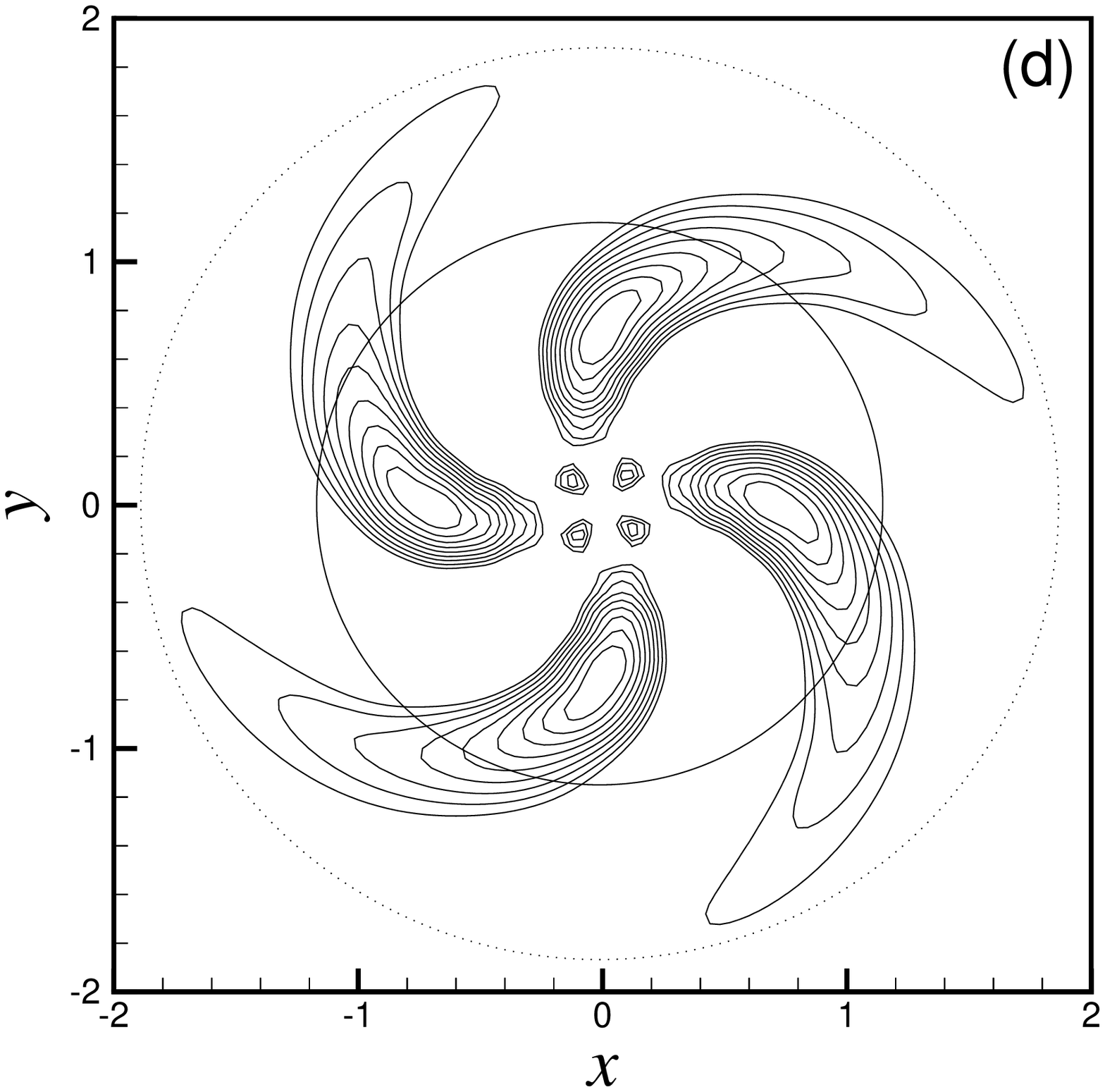}}}
\caption[]{Unstable modes of the cored exponential disc for $N$=6,
$\lambda=R_C/R_D$=0.625 and $G\Sigma_0 R_D/v_0^2$=0.34. Contours
show the positive surface density perturbations. The contour levels
are equally spaced from 10 to 90 percent of the maximum.
Solid and dotted circles mark the CR and OLR circles respectively.
(a) The fundamental $m=2$ mode. (b) The secondary $m=2$ (spiral) mode.
(c) $m=3$. (d) $m=4$.
}
\label{fig:fig2}
\end{figure*}

\section{The analytical model\label{sec-ana}}

We study the stability properties of a razor-thin disc
that has an exponential density distribution with a central core
of radius $R_C$:
\bqn
\Sigma_D(R) &=& \Sigma_0 \exp
\left(-\lambda\sqrt{1+R^2/R_C^2}\right ),\qquad \lambda=\frac{R_C}{R_D}
\label{eq::dens-disc}
\eqn
We assume that the disc is in equilibrium with the total
gravitational potential of the galaxy represented by the
logarithmic law
\bqn
\Phi_0(R) &=&  v_0^2 \ln\sqrt{1+R^2/R_{C}^2}. \label{eq::pot-disc}
\eqn
Here $\Phi_0(R)$ is the total potential of the disc and halo
components. The disc rotation curve is given by the equation:
\bq
v_{\rm rot}=\sqrt{R{{\rm d}\Phi_0 \over {\rm d}R}}={v_0 R\over
\sqrt{R_{C}^2+R^2}}. \label{eq::rot-curve}
\eq
Equation (\ref{eq::rot-curve}) gives the rotational velocity of
a collisionless disc if the velocity dispersion of the disc is equal
to zero. In a disc with a non-zero velocity dispersion, the mean rotational
velocity $\langle v_\varphi \rangle$
differs from the simple law given by equation (\ref{eq::rot-curve})
due to a collisionless `pressure' that influences the rotation of the
disc.

The density distribution given by equation (\ref{eq::dens-disc})
is characterized by three parameters, namely the central surface
density $\Sigma_0\exp(-\lambda)$, the radial scale length of the disc density
distribution $R_{D}$ (through $\lambda$), and the core radius $R_{C}$.
We use the core
radius $R_{C}$ and the asymptotic circular velocity $v_0$ as units
to normalize the problem. The time unit is then determined by the
ratio $R_{C}/v_0$. The dimensionless parameter
$S_0=G\Sigma_0 R_{D}/v_0^2$ gives the ratio of the mass of the disc
to the total mass of the model within the radius  $R \approx 5R_{C}$.

Following JH, we assume the distribution function of the disc
particles in the form:
\bqn
f(E,L) &=&
\Sigma_0 \sum_{n=0}^{N} {N\choose n}
\left({L \over R_C}\right)^{2n}g_n(E), \label{eq:DF-equilibrium}\\
g_n(E) &=& {(-1)^{n+1}
\over 2^n \sqrt{\pi} \Gamma (n+1/2)} \nonumber \\
&{}& \times {{\rm d}^{n+1} \over {\rm d}E^{n+1}}
\left ( e^{-2NE/v_0^2} e^{-\lambda\, e^{E/v_0^2}} \right ).
\label{eq:gn-equilibrium}
\eqn
Here $E$ and $L$ are the total energy and the angular momentum
of an individual star as
\bq
E=\frac 12 \left ( v_R^2+v_{\varphi}^2 \right )+\Phi_0(R),~~L=Rv_{\varphi}.
\eq
The family of distribution functions given by equations
(\ref{eq:DF-equilibrium}) and (\ref{eq:gn-equilibrium})
depends on an integer parameter $N$, that controls the fraction
of near-circular orbits, and thus controls the disc velocity
dispersion. The radial and azimuthal velocity dispersions
determined as
\bq
C_{R}=\sqrt{\langle v_R^2 \rangle},~~
C_{\varphi}=\sqrt{\langle v_\varphi^2\rangle-\langle
v_\varphi \rangle^2}
\eq
are small for relatively high values of $N$.
We select a moderately cold disc characterized by
the parameters $N=6$, $\lambda=R_C/R_D$=0.625 and $S_0$=0.34. The upper
panel of Figure \ref{fig:fig1} shows the rotation curve $v_{\rm rot}$
determined by equation (\ref{eq::rot-curve}), the mean rotational velocity
$\langle v_\varphi \rangle$ and the normalized surface density
$e^{\lambda}\Sigma_D(R)/\Sigma_0$ for this model. The lower panel of
Figure \ref{fig:fig1} shows Toomre's $Q$-parameter together with
the radial and azimuthal velocity dispersions as a function of
the radial distance $R$. For the selected model, the $Q$-parameter
is greater than unity everywhere in the disc, and the model is stable
for $m$=0 perturbations.

In the linear regime, the spiral density perturbation can be
written in polar coordinates $(R,\varphi)$ as:
\bq \Sigma_{m}(R,\varphi,t) =\hat{A}_{m,0}(R)e^{{\rm
i}(m\varphi-\omega t)}, \label{eq-sigm} \eq
where
\bq
\omega =m\Omega_p+{\rm i}s,~~
\hat{A}_{m,0}(R)=A_m(R)e^{{\rm i}\theta _m(R)}.
\eq
Here $A_m(R)$, $\theta_m(R)$, $\Omega_{p}$ and $s$ are the amplitude,
phase, pattern speed and growth rate of the $m$th global mode,
respectively.

Figure \ref{fig:fig2} displays the results of the linear global modal
analysis of our model. We find that the collisionless disc is unstable
towards two, three and four-armed spirals. We have shown the contour
plots of the unstable eigenmodes, together with the positions of the
corotation (CR) and outer Lindblad (OLR) resonances. Pattern speeds
$\Omega_{p}$ and growth rates $s$ of all unstable modes are collected
in Table \ref{tab-modes}.

Figure \ref{fig:fig2}{\em a} shows the fastest growing $m$=2 bar-like
mode (2p in Table \ref{tab-modes}). The dimensionless pattern speed and
growth rate for this mode are $\Omega_p=0.768$ and $s=0.642$, respectively.
Linear analysis also reveals another unstable $m$=2 global mode
(Figure \ref{fig:fig2}{\em b}) that
has a pattern speed of $\Omega_p=0.443$ and a growth rate of $s=0.119$
(2s in Table \ref{tab-modes}). This secondary mode is more spatially
extended than the fundamental bar-mode, and occupies the disc
within a few core radii. The spatial distribution of amplitude of the
bar-mode has a single maximum located approximately at a distance of
$R=0.25$ from the centre of the disc. The amplitude function of the
secondary $m$=2 mode has three maxima of comparable amplitudes shifted
approximately by 90 degrees with respect to each other. The secondary
$m$=2 mode has also been shown in Figure 9 of JH, and we reproduce it
here for completeness.

Figures \ref{fig:fig2}{\em c} and \ref{fig:fig2}{\em d} show the contour
plots for the unstable $m$=3 and $m$=4 modes found in the linear stability
analysis. These modes occupy approximately the same region of the disc as
the secondary $m$=2 mode does. These modes, however, grow faster than
the secondary $m$=2 mode (see Table \ref{tab-modes}). Therefore, the
dynamics of perturbations in the outer regions of the disc is governed
mostly by the $m$=3 and $m$=4 perturbations. The patterns of $m=3$ and
$m=4$ instabilities freely extend up to the outer Lindblad radius.
This shows that the corotation resonance has no influence on these
modes.

\section{Numerical simulations\label{sec-num}}

\subsection{Code and data analysis\label{sec-nbody}}

Our $N$-body simulations are carried out with the help of the {\sc Superbox}
code (Fellhauer et al. \cite{fel00}). This is a highly efficient
particle-mesh-code with nested and comoving grids, based on a
leap-frog scheme with second order force calculation. Nested grids,
comoving with the center of mass, allow us to achieve high resolution
in the central parts of the collisionless systems. The code has been
successfully applied to study mergers of galaxies
(Madejsky \& Bien \cite{mad93}), galaxy-satellite disruption
(Klessen \& Kroupa \cite{kle98}) and the orbital decay of satellite
galaxies (Pe\~narrubia et al. \cite{pen02},
Just \& Pe\~narrubia \cite{jus05}). It was also used to study orbital
evolution of a supermassive black hole in a galactic nucleus
(Spinnato et al. \cite{spi03}).

We use {\sc Superbox} to simulate the dynamics of perturbations in a
collisionless disc, which is in rotational equilibrium under the
influence of its self-gravity and the gravitational potential of an
external rigid halo. The code is three dimensional. To study the
dynamics of a two-dimensional gravitating disc, we confine the
initial distribution of particles to the $(x,y)$-plane and set
vertical velocities of the particles equal to zero. The dynamics of
the model was simulated using different sets of the numerical
parameters (i.e., number of particles, grid resolution, etc.) listed in
Table \ref{tab-sim}. The radial extent of the inner, intermediate
and the outer grid zones are determined by the parameters $R_{\rm i}$,
$R_{\rm m}$ and $R_{\rm o}$, and the spatial resolution in
each grid zone is determined for the inner grid zone by the parameter $d_{\rm i}$.

The {\sc Superbox} has a fixed time step, so
to resolve the motion of the particles in the central regions of the
disc, we use a time step of ${\rm d}t=0.05$\,Myr. This is less than a
typical crossing time of a grid cell in the inner regions of the
disc ($\approx0.1$\,Myr).
\begin{table*}[t]
\begin{tabular}{lrlrrlrrrrrr} \hline
Model & $dim$& $\lambda$&$R_{max}$&
  $\Sigma_0$  & dt& $N$& grids&
 $R_{\rm o}$& $R_{\rm m}$& $R_{\rm i}$ &$d_{i}$\\
& & &kpc& $M_{\odot}/pc^{2}$&Myr&$10^6$&&kpc&kpc&kpc&pc\\
\hline\hline
 A0& 2D & 1.0 & 16.0 & 500 & 0.05 &   $4$ & $64^3$ & 22.0 & 5.0 & 1.0 & 33.3 \\
 A1& 2D & 1.0 & 16.0 & 500 & 0.05 &  $4$ & $128^3$ & 22.0 & 5.0 & 1.0 & 16.1 \\
 A2& 2D & 1.0 & 16.0 & 500 & 0.05 &   $13$ & $128^3$ & 22.0 & 5.0 & 1.0 & 16.1 \\
 A3& 3D & 1.0 & 16.0 & 500 & 0.05 &   $13$ & $128^3$ & 22.0 & 5.0 & 1.0 & 16.1 \\
\hline
 B0& 2D & 0.625 & 16.0 & 1233 & 0.05 &  $4$ & $64^3$ & 17.0 & 4.0 & 1.0 & 33.3 \\
 B1& 2D & 0.625 & 16.0 & 1233 & 0.05 &  $13$ & $128^3$ & 17.0 & 4.0 & 1.0 & 16.1 \\
 B2& 2D & 0.625 & 16.0 & 1233 & 0.05 &  $13$ & $128^3$ & 17.0 & 4.0 & 1.0 & 16.1 \\
 B3& 2D & 0.625 & 16.0 & 1233 & 0.05 &  $13$ & $128^3$ & 17.0 & 4.0 & 1.0 & 16.1 \\
 B4& 2D & 0.625 & 16.0 & 1233 & 0.05   & $13$ & $128^3$ & 17.0 & 4.0 & 1.0 & 16.1 \\
 B5& 2D & 0.625 & 14.0 & 1233 & 0.025 & $13$ & $128^3$ & 17.0 & 4.0 & 1.0 & 16.1 \\
 B6& 2D & 0.625 & 12.0 & 1233 & 0.05 &   $20$ & $128^3$ & 17.0 & 4.0 & 1.0 & 16.1 \\
 B7& 2D & 0.625 & 13.0 & 1233 & 0.05   & $40$ & $128^3$ & 17.0 & 4.0 & 1.0 & 16.1 \\
 B8& 2D & 0.625 & 13.0 & 1233 & 0.05  & $13$ & $256^3$ & 17.0 & 4.0 & 1.0 & 7.9 \\
 B9& 3D & 0.625 & 16.0 & 1233 & 0.05   & $13$ & $128^3$ & 17.0 & 4.0 & 1.0 & 16.1 \\
 B10&2D & 0.625 & 12.0 & 1233 & 0.02   & $3$ & $128^3$ & 17.0 & 4.0 & 1.0 & 16.1 \\
 B11&2D & 0.625 & 12.0 & 1233 & 0.02   & $13$ & $128^3$ & 17.0 & 4.0 & 1.0 & 16.1 \\
 B12&2D & 0.625 & 12.0 & 1233 & 0.02   & $30$ & $128^3$ & 17.0 & 4.0 & 1.0 & 16.1 \\
\hline
\end{tabular}
\caption{Parameters of the different models. All models have the
same circular velocity $v_0=220\,km/s$ and disc scale length
$R_{D}=3\,kpc$. Model A with $\lambda=R_{C}/R_{D}=1$ and $S_0=0.133$
is the stable reference model. Model B with $\lambda=0.625$ and
a larger disc mass $S_0=0.329$ corresponds to the unstable model in
\S\ref{sec-ana}. 
$dim$  and $R_{max}$ are the dimension and cutoff radius of the
disc. $\Sigma_0$ determines the surface density of the disc (see Eq.
\ref{eq::dens-disc}) and $dt$ is the time step of the models.
$N$ gives the number of
particles, the next column the number of grid cells in each grid.
$R_{\rm i}$, $R_{\rm m}$ and $R_{\rm o}$ are the radii of the inner,
middle and outer grid. $d_{i}$ gives the cell size of the inner grid
cells.} \label{tab-sim}
\end{table*}
%

\begin{figure}[t]
\centerline{
  \resizebox{0.98\hsize}{!}{\includegraphics[angle=270]{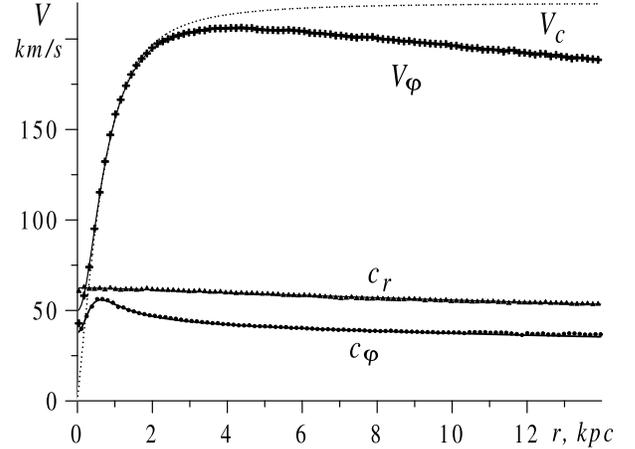}}
  }
\caption[]{The rotational velocity and velocity dispersions of the
disc as a function of the radial distance $R$. Solid lines are the
disk equilibrium profiles calculated with help of the distribution
function (equation \ref{eq:DF-equilibrium}).
The points correspond to the $N$-body
realizations of the initial equilibrium. Dotted line shows the
rotation curve $v_0(R)$ of equation (\ref{eq::rot-curve}). }
\label{fig:fig3}
\end{figure}
\begin{figure*}[t]
\centerline{
 \resizebox{0.9\hsize}{!}{\includegraphics[angle=0]{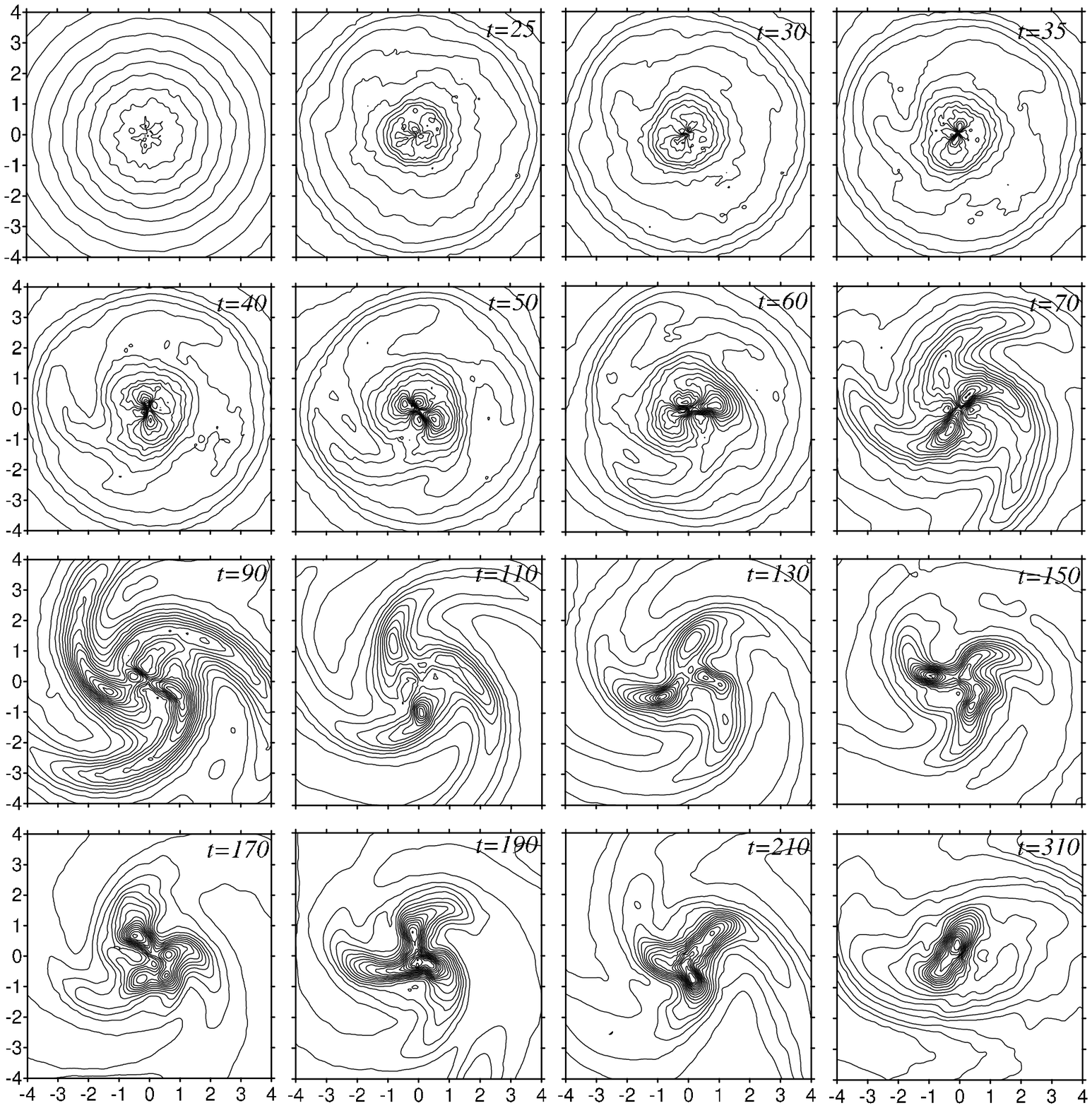}}
  }\vskip -0.05\hsize
\caption[]{Temporal evolution of the surface density distribution
for an unstable disc (model shown in Fig. 2). A
rapidly growing rotating bar-mode emerges in the central
regions of the disc at early stages of disc evolution.
At later times, the more slowly growing three-armed spiral
determines the dynamics of perturbations in the outer regions of the
disc.  The time $t$ is given in Myr.}
\label{fig:fig4}
\end{figure*}

The disc is built with equal mass particles. To construct a 2D
collisionless model in dynamical equilibrium, we use the distribution
function given by equation (\ref{eq:DF-equilibrium}), and set
velocities of the particles in $z$-direction and their
$z$-coordinates equal to zero ($z_{j}=v_{z,j}=0$).
The gravitational potential of a rigid halo is assumed to be
spherical and is calculated by subtracting
the disc potential $\Phi_D$ from the total
potential given by the equation (\ref{eq::pot-disc}).

To determine the macroscopic characteristics 
of the disc, as a function of radius, such as surface density $\Sigma(R)$, 
disc rotation $\langle v_{\varphi} \rangle$ and the radial and 
azimuthal velocity dispersions $C_{R}$ and $C_{\varphi}$,
we divide the disc into $n_{r}=100$ rings of equal width and calculate
the macroscopic values in each ring. Figure \ref{fig:fig3}
shows the $N$-body realization of the initial equilibrium profiles
of the disc. In this Figure, the solid lines show the the rotation
velocity and the velocity dispersions of the disc,
calculated with equation (\ref{eq:DF-equilibrium}), and
the points are the
$N$-body realization of disc equilibrium.
As it can be judged from Figure \ref{fig:fig3}, there is satisfactory
agreement between the theoretical model and its $N$-body
realization except for the very central regions of the disc.

To quantify the growth rates and the pattern speeds of the unstable
modes, we calculate the Fourier components of the perturbed density
for azimuthal wavenumbers $m=1,2,\cdots,6$. With point-like
particles, the Fourier components are given by the real part of the
expression
\bq \tilde{\Sigma}_{m}=\tilde{A}_{m}(R_{i})e^{i m\varphi}
=\frac{M_i}{S_{i}}\sum_{j=1}^{N_{i}}e^{i m(\varphi-\varphi_{j})},
\eq
where $S_i$ is the area of the $i$th ring element corresponding to
$R_i$, $N_i$ is the number of particles in the $i$th ring, and $M_i$
is the total mass of particles that lie on $S_i$. By defining the
amplitude $A_{m}(R_{i})=\left|\tilde{A}_{m}(R_{i})\right|$ we get
\bq
\Sigma_{m}=A_{m}(R_{i})\cos \left( m\varphi-\theta_{m} \right).
\label{eq-ampl}
\eq
Generally, the amplitude $A_{m}(R,t)$ and the phase $\theta_{m}(R,t)$
of the modes depend on time and radius. By measuring these quantities,
we can estimate the growth rate $s$ and the pattern speed $\Omega_{p}$
of the modes by fitting the perturbed quantities to the expressions
\bq
A_{m}(t)\to A_{m,0}e^{st},~~
\theta_{m}(t)\to\theta_{m,0}+m\Omega_{p}t.
\eq
If the growth rate and the pattern speed of a Fourier component
are independent of radius, the perturbation is dominated by an
eigenmode.

To test the influence of the resolution effects and to
determine the long-term behaviour caused by numerical errors,
we simulated the dynamics of a stable model with a low disc mass
(model A in Table \ref{tab-sim}). For this model, the number of
particles as well as the grid resolution has been varied. We find
that after adapting the particle distribution to a grid-based
potential the disc stays in equilibrium. After $3\,Gyr$, the artificial
heating caused by the particle-mesh scattering is lower than 5\%
of the initial velocity dispersion which shows that
{\sc Superbox} is intrinsically collisionless.

To test the dynamics of a three-dimensional stellar disc, we expanded
model A2 in the vertical direction to an isothermal slab with the 
scale height of $h=100$\,pc (model A3). The result of this exercise 
is that the dynamical equilibrium is not affected by the thickening 
and that the stability of the disc is not destroyed.

\subsection{The unstable model}

In this section we discuss properties of the unstable modes
found in numerical simulations,
and compare the results to those obtained in a linear analysis.
We find that the low-resolution model B0 leads to a significant discrepancy
between N-body simulations and analytical predictions.
We therefore discuss results obtained for the higher-resolution model B1
which is built with 13 million particles. Models B2-B8 differ from the 
model B1 in the numerical parameters, namely the outer 
cutoff radius, the time step, the number of particles, and the grid 
resolution, and serve to illustrate that parameter variations do not affect 
the results of N-body simulations for our fiducial model.
Model B9 follows the disc evolution three times longer than B1 with the
same parameters to investigate the nonlinear evolution of the unstable modes.
Models B10-B12 are a series, where only the number of particles is changed for
the determination of the N dependence of pattern speeds and growth rates.

Figure \ref{fig:fig4} shows a time sequence for the contour plots of the
perturbed disc surface density plotted up to the nonlinear saturation phase.
Linear analysis of our model predicts the existence of a few
simultaneously growing unstable modes. The results of N-body
simulations are in qualitative agreement with the linear analysis predictions.
As expected, the early stages of disc evolution
are governed by the fastest growing $m$=2 global
bar-mode, developing in disc central regions.
This mode is seen in Figure \ref{fig:fig4}
starting from 25 Myr.  By 60 Myrs,
the central bar-mode saturates, and the perturbations continue to grow
in the outer disc regions. The fastest growing mode is confined to the 
central kiloparsec of the disc, and the dynamics of the outer disc 
regions is governed by more slowly growing spirals. The disc dynamics 
in outer regions is determined by a superposition of slowly growing
two-armed, three-armed and four-armed spirals. These modes
are shown in Figure \ref{fig:fig2}. $N$-body simulations
confirm this finding.
 
\begin{figure*}[!t]
\centerline{
  \resizebox{0.4\hsize}{!}{\includegraphics[angle=0]{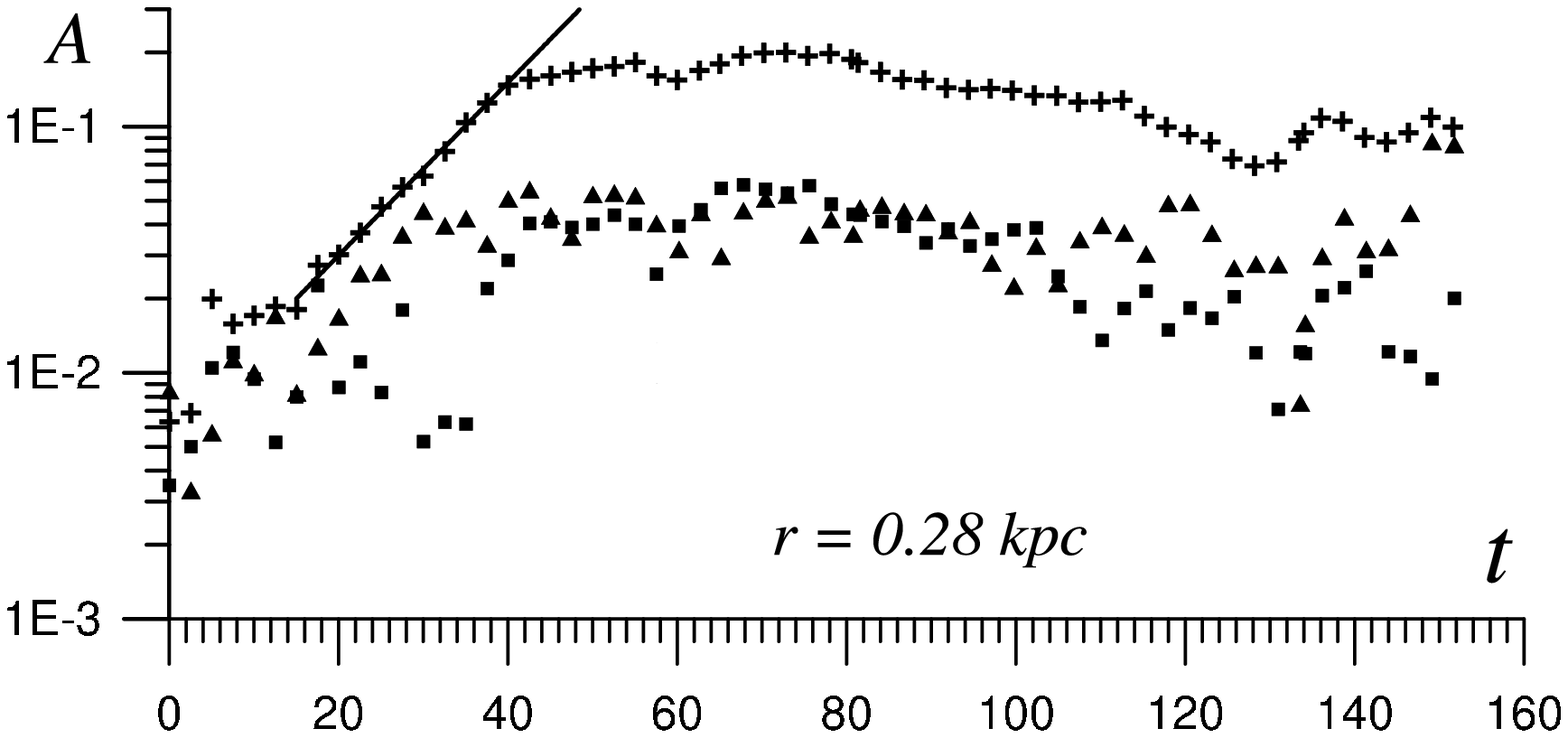}}
  \resizebox{0.4\hsize}{!}{\includegraphics[angle=0]{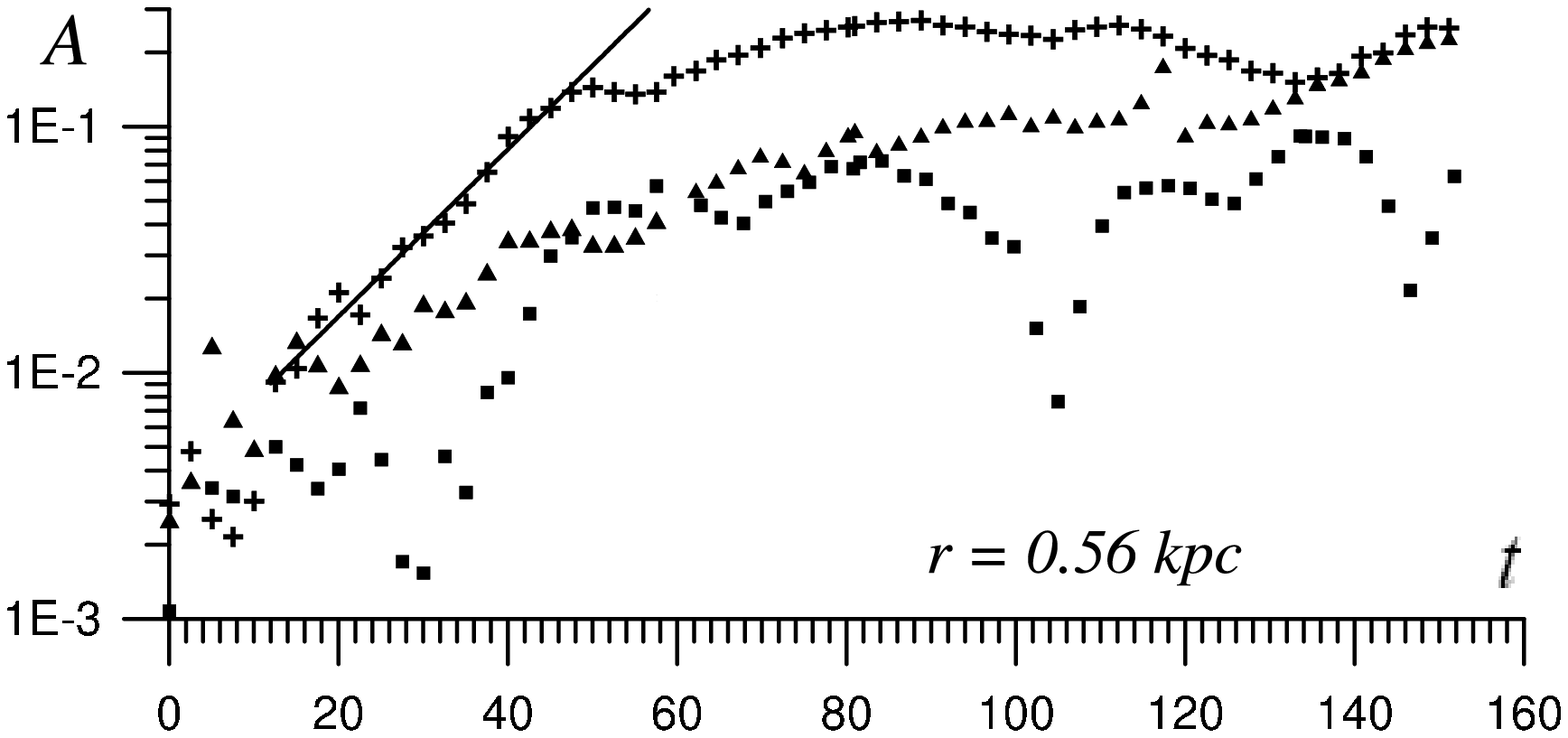}}
  }
\centerline{
  \resizebox{0.4\hsize}{!}{\includegraphics[angle=0]{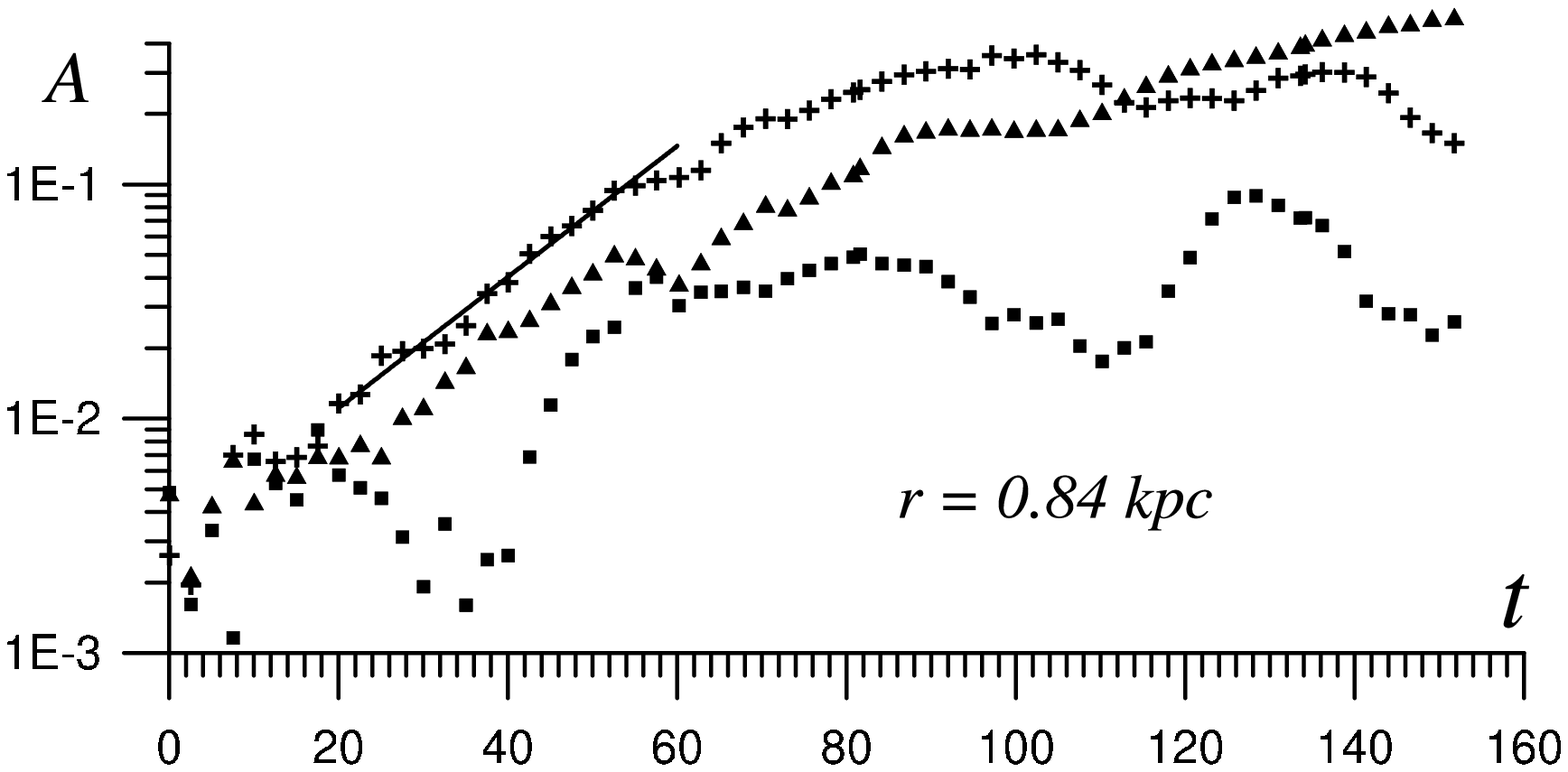}}
  \resizebox{0.4\hsize}{!}{\includegraphics[angle=0]{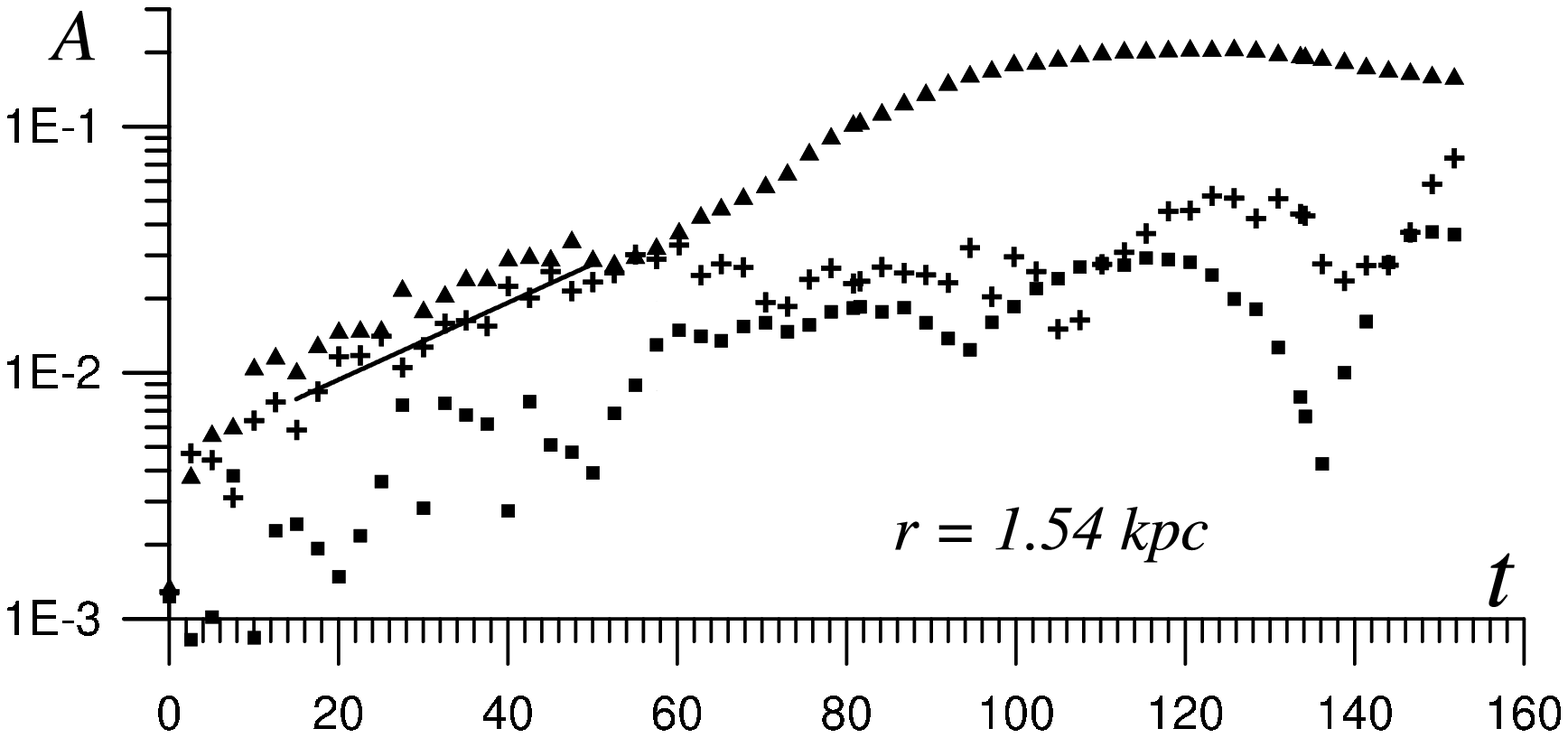}}
  }
\centerline{
  \resizebox{0.4\hsize}{!}{\includegraphics[angle=0]{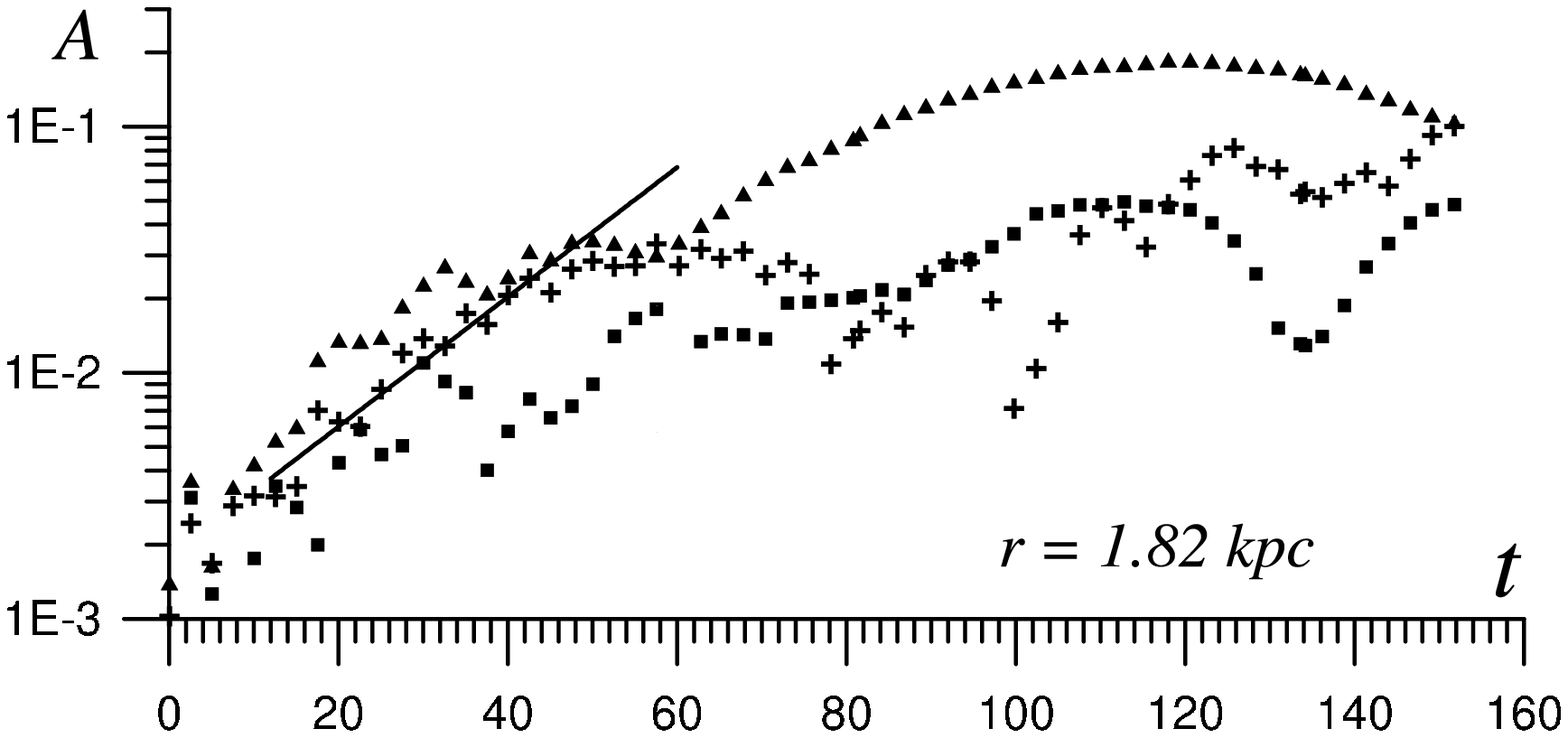}}
  \resizebox{0.4\hsize}{!}{\includegraphics[angle=0]{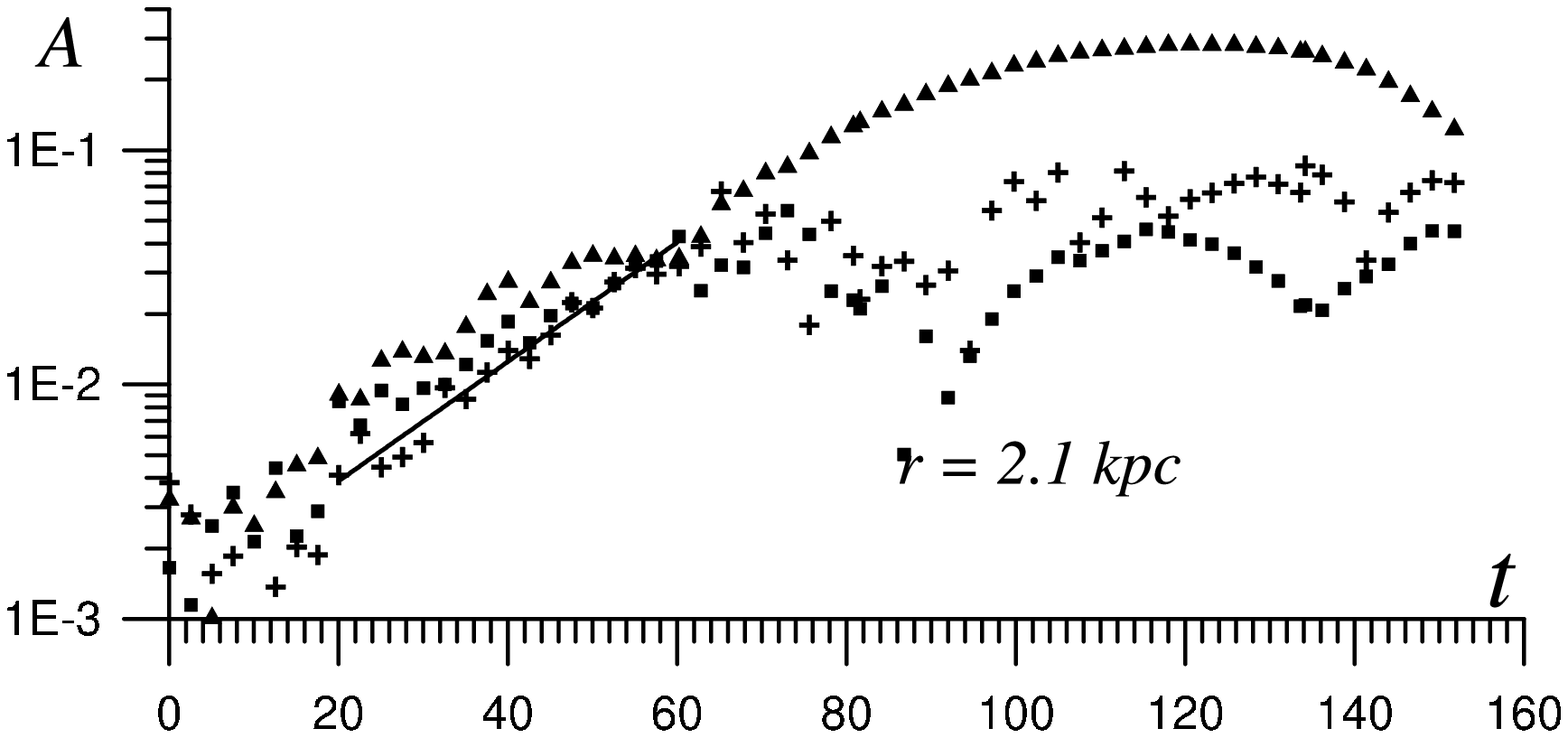}}
  }
  \vskip -0.07\hsize
\centerline{
  \resizebox{0.68\hsize}{!}{\includegraphics[angle=270]{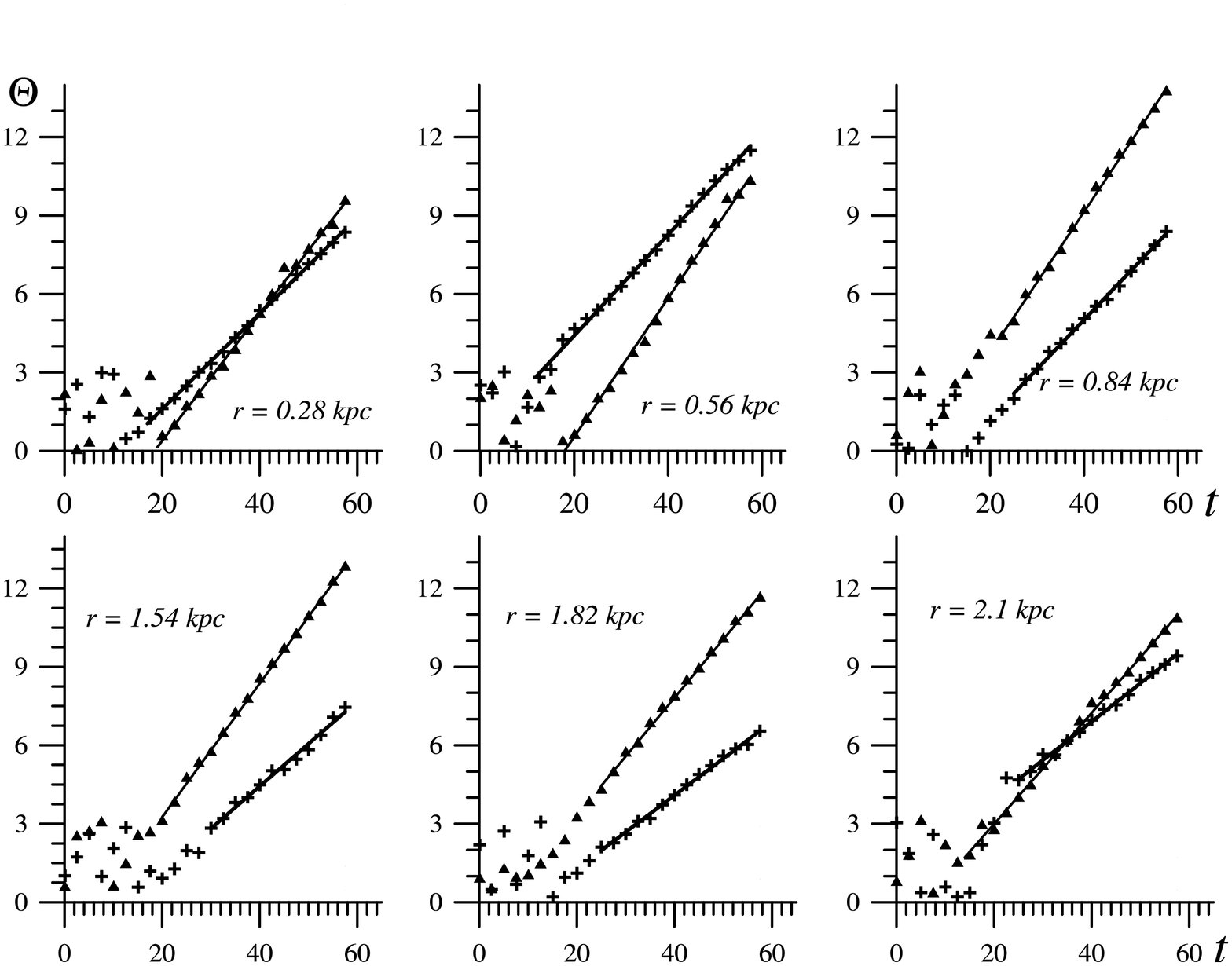}}
  }
\caption[]{ The dependence on time of $m=$2 (crosses), $m=$3
(triangles), and $m=$4 (squares) amplitudes of the global modes
$A$, measured at different radii. In the central regions of the
disk the fastest growing is the $m=$2 global bar-mode. At the disc
periphery the three armed perturbation prevails over other
competitors. The time evolution for the phases of $m=$2 (crosses)
and $m=$3 (triangles) Fourier components $\Theta$ is shown in the
lower frames. } \label{fig:fig5}
\end{figure*}

For quantitative comparison
of growing global modes found in $N$-body
simulations with the analytically predicted results, we calculate a Fourier
decomposition of the density perturbations. This allows us to determine
the amplitude $A_{m}(r,t)$ and phase $\Theta_{m}(r,t)$ of an
$m$-armed spiral mode as a function of time and of the disc radius, and to
determine the growth rate and the pattern speed of each Fourier component as
a function of radius. Figure \ref{fig:fig5} shows the amplitude
and the phase of Fourier modes $m=2,3,4$ calculated at different radii
in the disc. In the central regions, the dynamics of perturbations
is governed by the fastest growing $m$=2 mode, which saturates at 
about 40 Myr. In the outer disc regions, the dynamics is dictated 
by the $m$=3 perturbation prevailing over its $m$=2, and $m$=4 
competitors. As can be judged from Figure \ref{fig:fig5}, the spiral 
growth rates in the linear stage and their pattern speeds are 
fairly independent of radius, which indicates that perturbations 
are indeed the  growing global modes.
The growth rates $s$ and pattern speeds $\Omega_p$ for the primary
mode $m=2$ do not change value at the transition from the linear stage
to the essential nonlinear stage ($t>>1/s$).

\begin{figure}
\centerline{
  \resizebox{0.8\hsize}{!}{\includegraphics[angle=0]{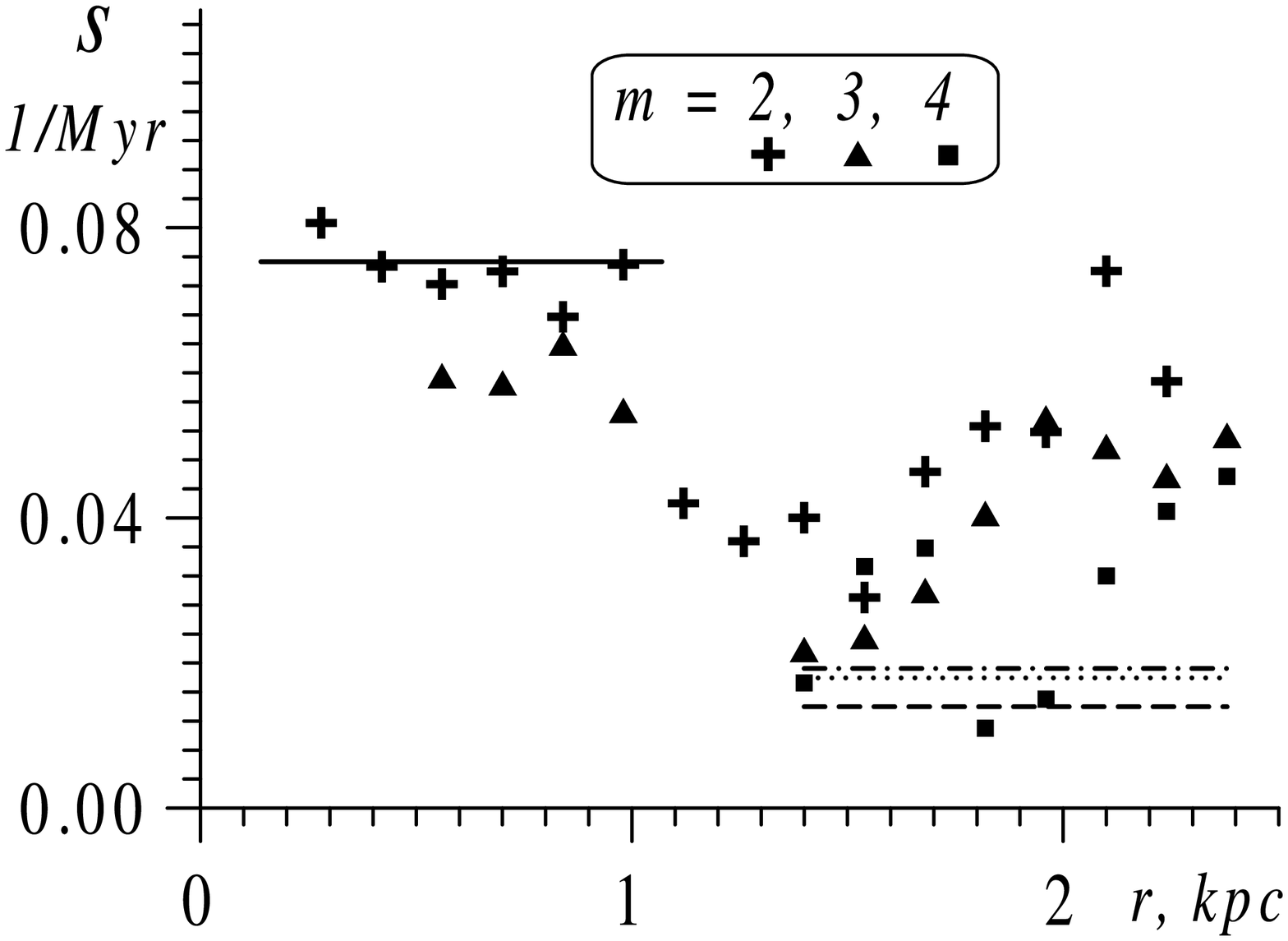}}
  }
\centerline{
  \resizebox{0.8\hsize}{!}{\includegraphics[angle=0]{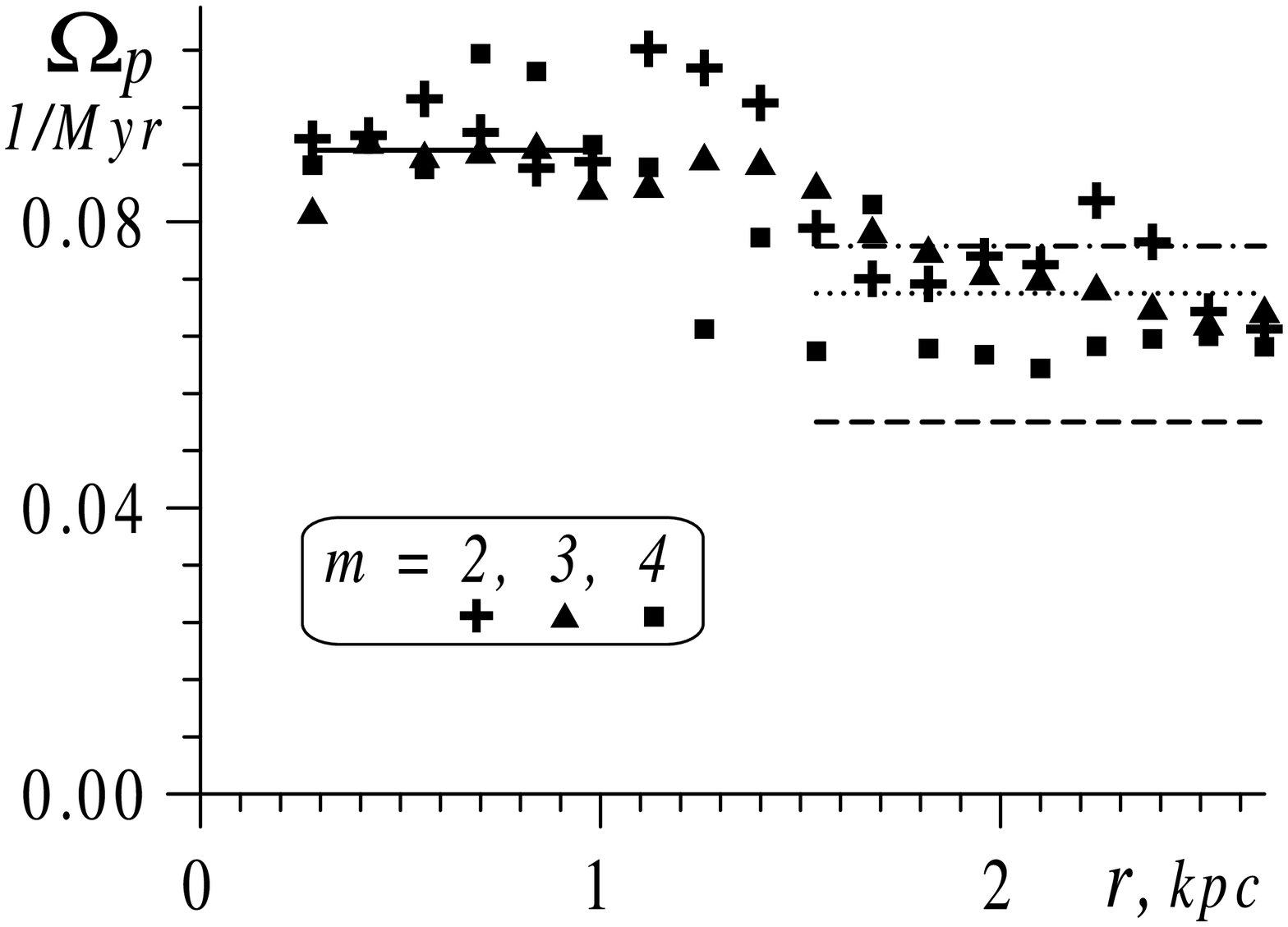}}
  }
\centerline{
  \resizebox{0.8\hsize}{!}{\includegraphics[angle=0]{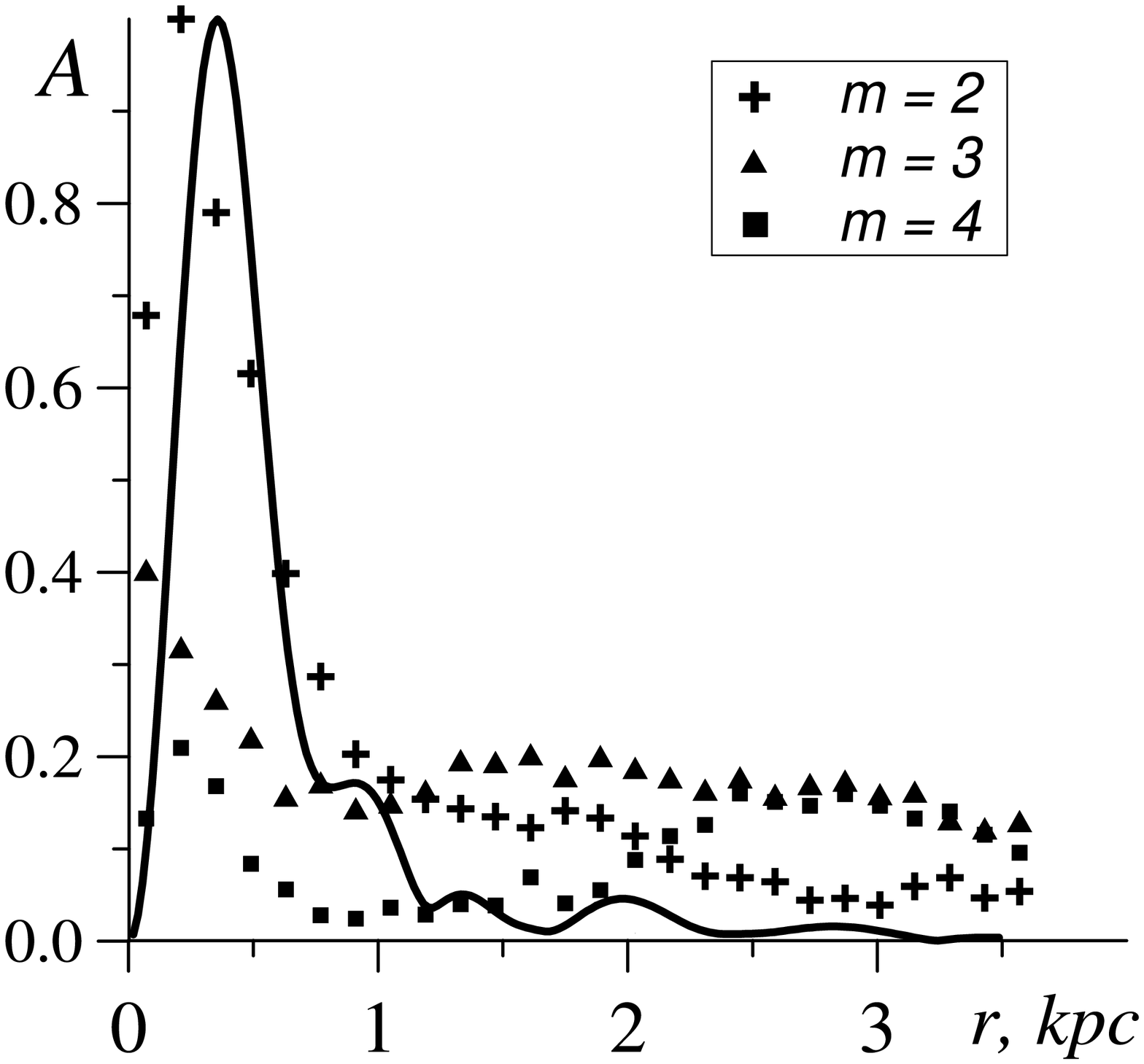}}
  }
 \vskip -0.08\hsize
  \caption[]{
{it Upper and middle panels}: growth rates and pattern speeds for $m$=2,
$m$=3 and $m$=4 Fourier components measured at different radii. In the
central region of the disc, the growth rate and pattern speed of the $m$=2
Fourier component agree with the linear analysis prediction of the bar
mode (solid lines). Analytic predictions of the growth rates and pattern
speeds are indicated at larger radii for $m$=2 (dashed lines), $m$=3
(dotted lines) and $m$=4 (dot-dashed lines) Fourier components.
{\it Lower panel:} Points - the amplitudes of $m$=2, 3 and 4 -
Fourier components as a function of radius taken from $N$-body
simulations at time 40 Myr. The amplitude profile for the most
unstable $m$=2 central bar-mode agrees qualitatively well with the
linear analysis (solid line). }
\label{fig:fig6}
\end{figure}

This is additionally illustrated in Figure \ref{fig:fig6} showing
the radial dependence of the growth rates, pattern speeds (upper
frames), and the amplitudes of $m$=2, 3, 4 Fourier components as a
function of radius. The parameters have been calculated from
$N$-body simulations at time 40 Myr. The fastest growing
bar-per\-turbation grows as a whole in the central regions of the
disc, and rotates with fairly constant angular velocity. The
measured values for the growth rate and for the pattern speed for
this mode agree within errors of measurement with the linear
analysis. The pattern speed of the bar-mode of $\Omega_p=0.78$
agrees with the theoretical value of $\Omega_p=0.77$, and the
growth rates found in both linear analysis and $N$-body
simulations quantitatively agree as well. The amplitude of the
bar-mode reaches its maximum at about 0.25 kpc (bottom frame of
Figure \ref{fig:fig6}), and and then decreases with radius. Such a
radial profile agrees qualitatively with the linear analysis
predictions (solid line). A discrepancy between the linear
analysis and numerical simulations in the outer regions of the
disc is explained by the presence of a secondary $m$=2 global
mode.

The agreement between growth rates and pattern speeds for
$m$=3, and $m$=4 modes measured in N-body simulations and those
determined in linear analyses are less satisfactory.
While pattern speeds agree reasonably well, the growth rates
measured in N-body simulations show a considerable mismatch
with the values found in linear analyses. Being lower than
the growth rate of the fastest growing bar-mode, the growth rates
in N-body simulations are higher than the analytical
predictions. In part, the discrepancy can be related to the
still low resolution of our simulations which is not high enough to 
properly model multi-modal behaviour of the unstable disc.

\begin{figure}[t]
\centerline{
  \resizebox{0.8\hsize}{!}{\includegraphics[angle=0]{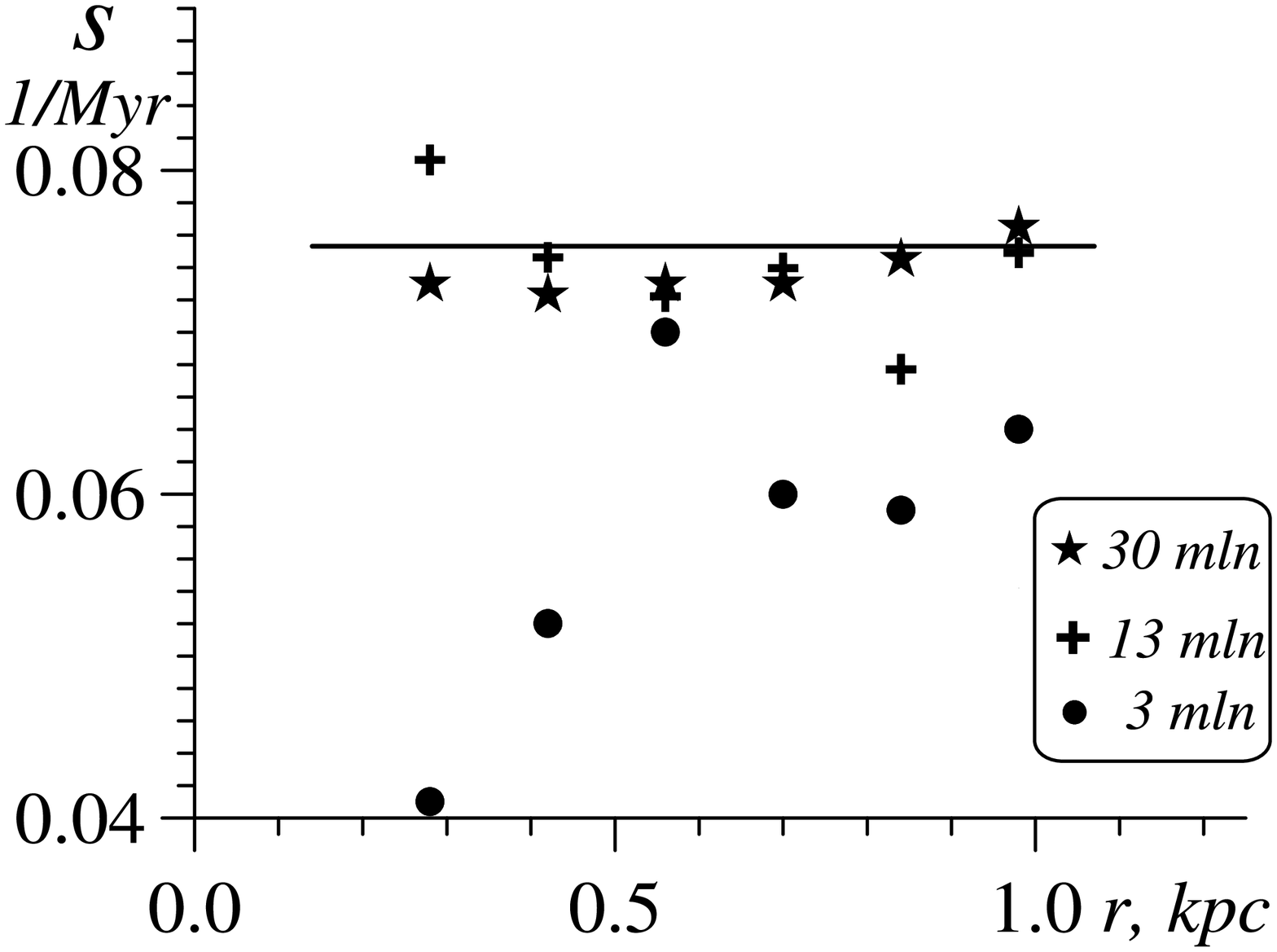}}
  }
\centerline{
  \resizebox{0.8\hsize}{!}{\includegraphics[angle=0]{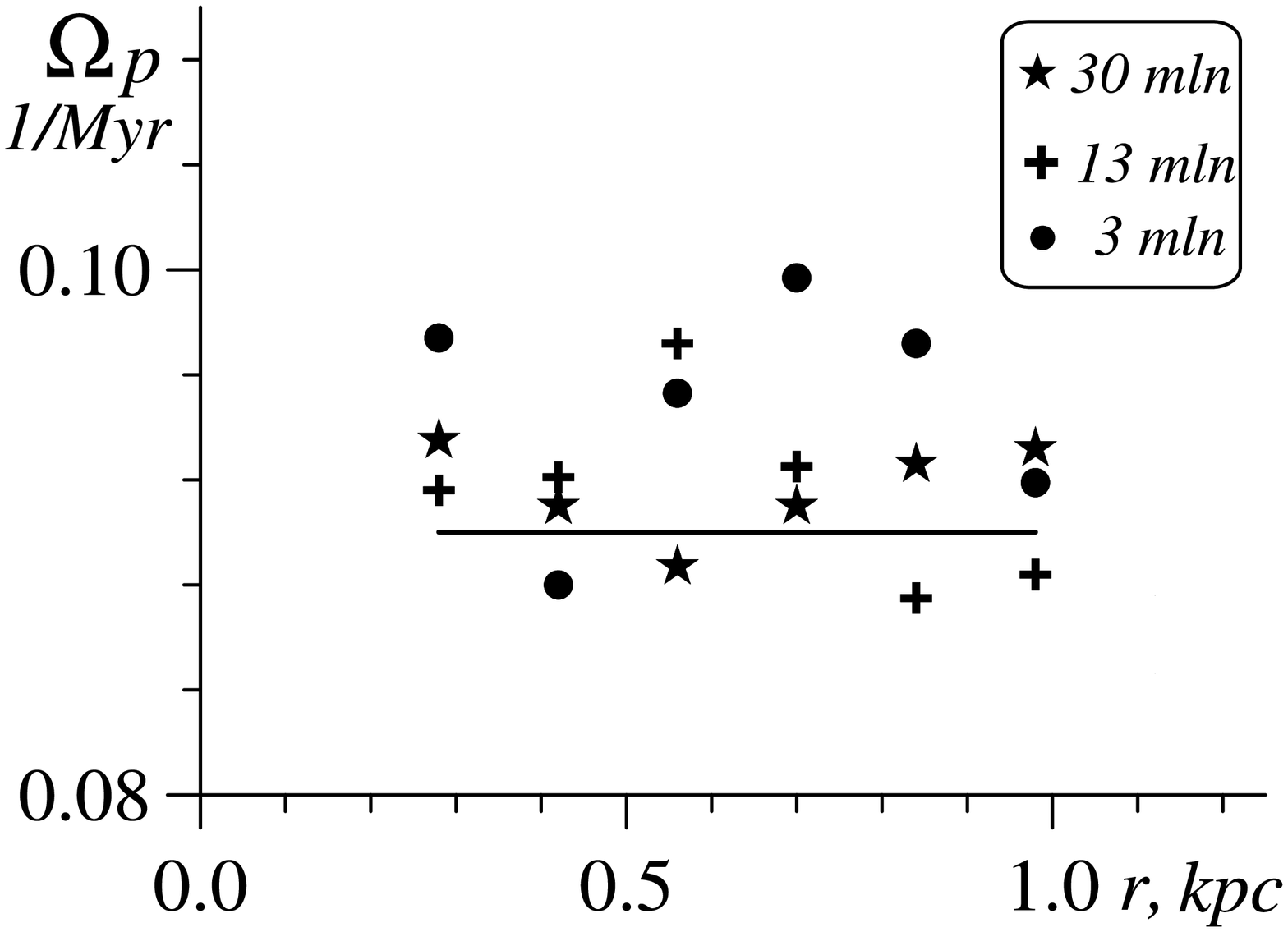}}
  }
\caption[]{
The growth rate and the pattern speed of the principal $m=$2 bar-mode
as a function of a number of particles in N-body simulations.
}
\label{fig:fig7}
\end{figure}

Figure \ref{fig:fig7} shows the dependence of the growth rate and
the pattern speed of the principal bar-mode as a function of the
number of particles used in N-body simulations. As can be seen from
Figure \ref{fig:fig7}, the parameters of the unstable mode depend
strongly on resolution, i.e., number of particles used in N-body
simulations. The experiment with 3 million particles gives a large
discrepancy between the predicted parameters of the unstable mode
(solid lines in Figure \ref{fig:fig7}) and the values measured in
N-body simulations. A simulation with 13 million particles gives
better agreement with theory, and one with 30 million particles
gives a satisfactory agreement. Additionally,
SUPERBOX achieves best resolution in the disc's central regions,
so the parameters of the centrally confined fastest growing bar-mode
are in better agreement with the analytical predictions compared
to the more slowly growing modes developing in the outer regions
of the disc. An increase of particle resolution in N-body simulations
should lead to a better agreement of the parameters of the secondary
unstable modes with theory.

The spatial appearance of the eigenmodes is
determined by the radial dependence of the amplitude and the phase
of each Fourier component.
\begin{table}[t]
\begin{tabular}{r|rrrr|rr} \hline
 mode & \multicolumn{4}{c}{analytical} &\multicolumn{2}{c}{simulations}\\
 $m$ & $\Omega_{p}$ & $s$ & $R_{\rm CR}$ & $R_{\rm OLR}$ &
$\langle \Omega_{p}\rangle$ & $\langle s\rangle$ \\
\hline
2p & 0.768 & 0.642 & 0.834 & 2.077 & 0.78 & 0.63 \\
2s & 0.443 & 0.119 & 2.042 & 3.775 & 0.61 & 0.41 \\
3  & 0.597 & 0.153 & 1.345 & 2.320 & 0.62 & 0.33 \\
4  & 0.653 & 0.164 & 1.159 & 1.880 & 0.57 & 0.24 \\
\hline
\end{tabular}
\caption{Growth rates and pattern speeds of the different modes
for model B12. The time unit is $R_{C}/v_0=8.52\,$Myr.}
\label{tab-modes}
\end{table}

The  $m$=2 Fourier component is a superposition of the primary
and of the secondary unstable modes. We could not achieve a reliable
decomposition of the two $m$=2 modes because of the noise and the
nonlinear effects. Nevertheless, it is possible to estimate which
of the two $m=$2 modes is dominant in a particular region of the disc.

\begin{figure*}[!t]
\centerline{
  \resizebox{0.8\hsize}{!}{\includegraphics[angle=0]{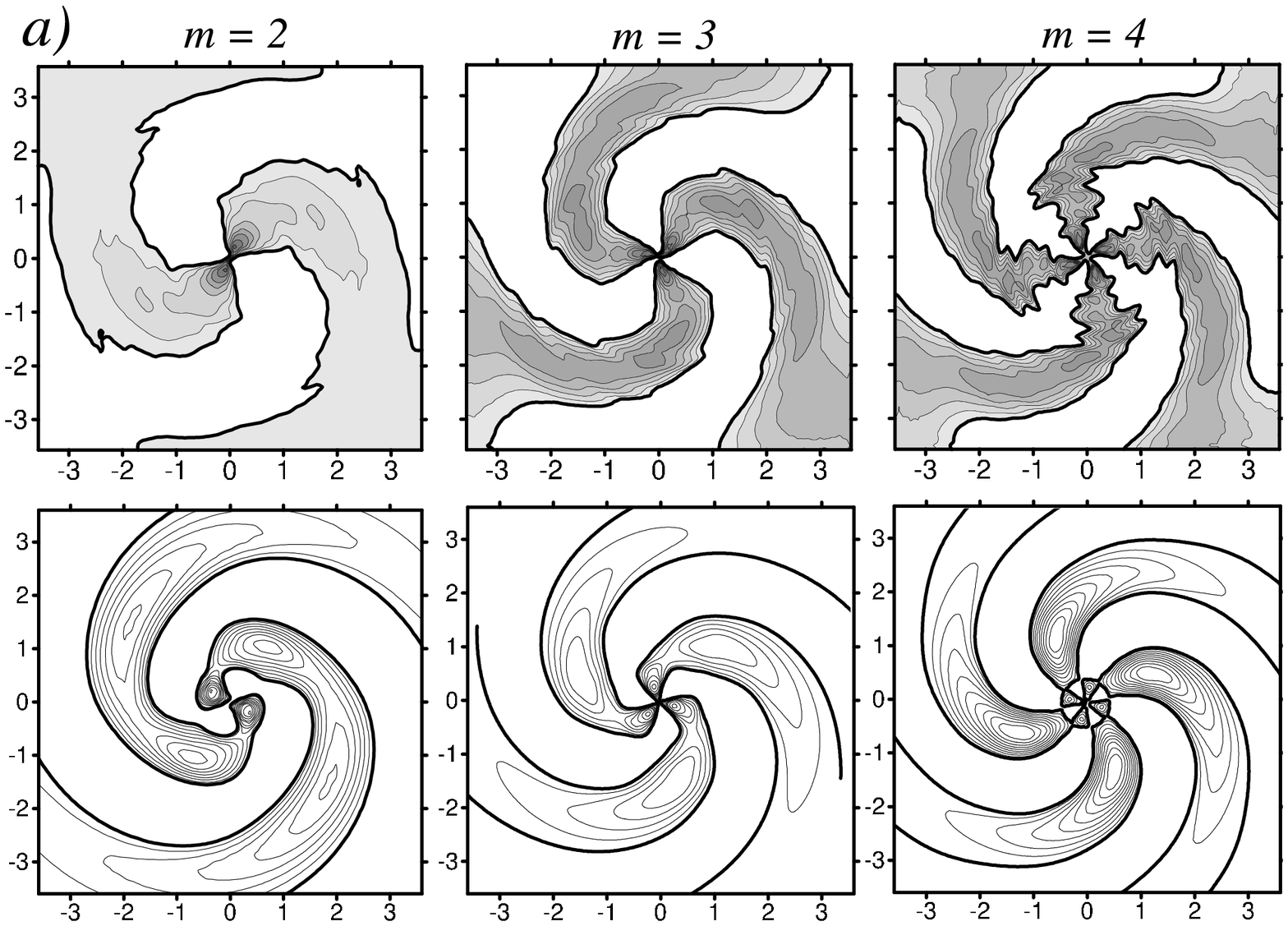}}
  }
\centerline{
  \resizebox{0.8\hsize}{!}{\includegraphics[angle=0]{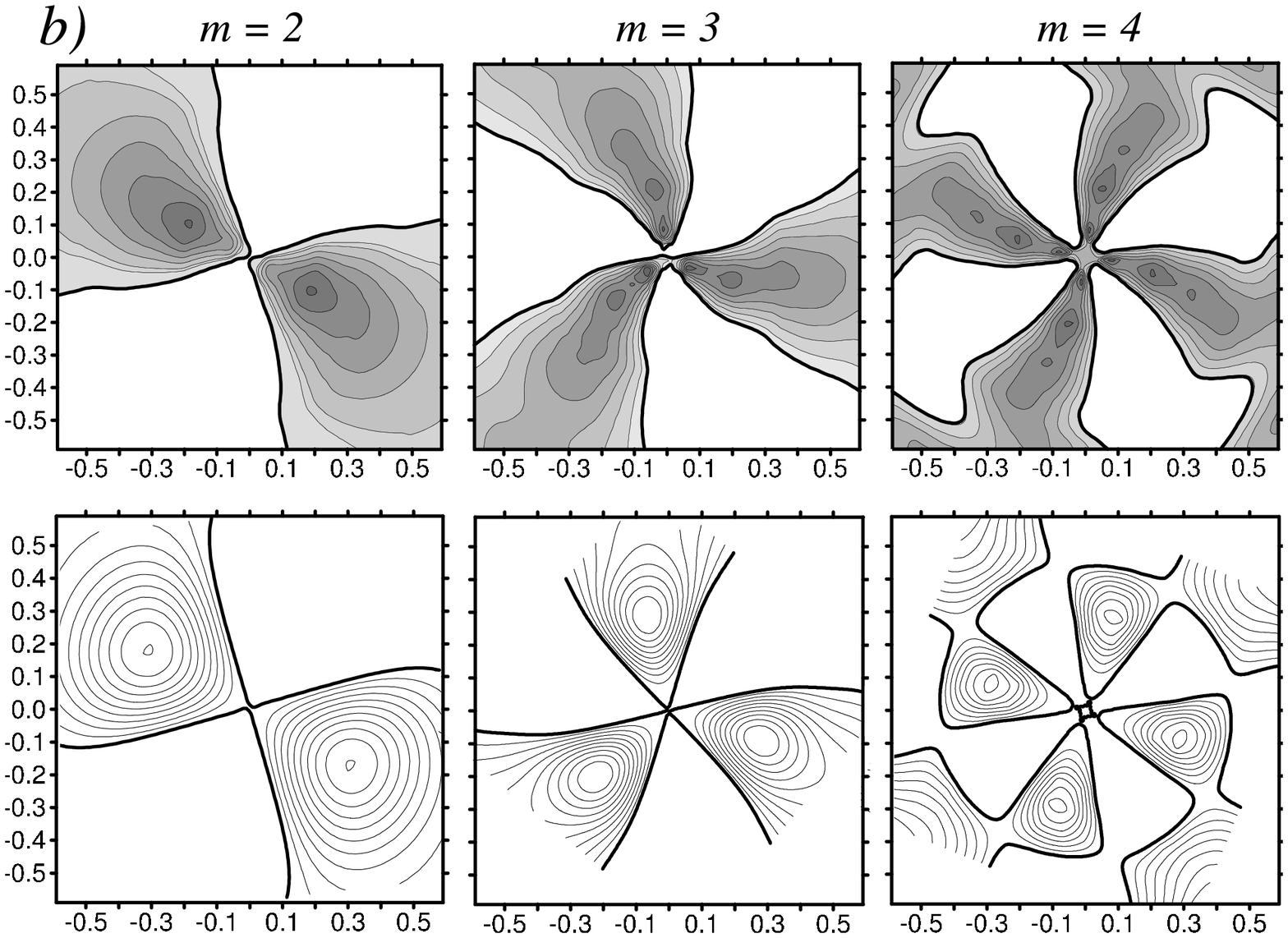}}
  }
\caption[]{a) Top frames - the surface density distribution for
$m$=2, 3 and 4 Fourier components determined from $N$-body
simulations as compared to the surface density contour plots for
the unstable modes determined from a linear stability analysis
(bottom frames). For $m$=2 the secondary mode is shown (lower
left).
 b) Density distributions within a central half-kiloparsec
for the Fourier components shown in Fig.~\ref{fig:fig8}a. Here the
primary $m$=2 mode is plotted at the lower left for comparison.
 } \label{fig:fig8}
\end{figure*}

The unstable modes developing in $N$-body simulations agree
qualitatively with the linear analysis not only in the growth rate
and the pattern speed, but also in their spatial appearance.
Figure \ref{fig:fig8} shows the contour plots of the surface
density for the $m$=2, 3 and 4 Fourier components as determined
from $N$-body simulations (top frames) compared to the
surface density contour plots for the unstable modes calculated by
a linear stability analysis. For comparison, only the
secondary $m$=2 linear mode is shown in the bottom left frame of
Figure \ref{fig:fig8}a. As one can see, the spatial range, and
winding of the unstable modes agree qualitatively in both $N$-body
and linear predictions.

Figure  \ref{fig:fig8}b shows a comparison of spatial
distributions of perturbations within the central half-kiloparsec
of the disc. The density distribution for the central bar-mode
agrees with the analytically predicted density distribution
plotted in the lower left panel. An agreement for density
distributions of $m$=3, and $m$=4 modes is also qualitatively
acceptable but the details are different because of the limited
resolution of the $N$-body model discussed above.

\section{Summary\label{sec-disc}}

We use high resolution $N$-body simulations to follow the dynamics
of growing spiral perturbations that develop in a collisionless disc
from the initial noise perturbations. At least ten million
particles with a minimum grid resolution of $128^3$ are needed to
reach a robust accuracy level so that the evolution of perturbations
is no longer dominated by noise.

Comparison of
$N$-body simulations with the results of linear stability analysis shows
an agreement between both approaches. The most unstable global
bar-mode developing from the noise perturbations has a
pattern speed and a growth rate as predicted by theory.
Other unstable modes (the primary $m$=3, $m$=4 modes, and a
weak secondary $m$=2 mode) qualitatively agree with
the theoretical results. However the pattern speeds and
growth rates of the secondary two-armed spiral, the
three-armed and four-armed spirals are in lesser agreement with
the theoretical values due to a still low resolution of the disc
dynamics in outer regions.

We have demonstrated that a fast particle-mesh code {\sc Superbox}
is useful to follow a detailed dynamics of a collisionless stellar
disc. Due to the large number of particles which can be used in
SUPERBOX simulations, the noise level, and the numerical heating can be
reduced to an insignificant level which allows us to follow the complex
dynamics of a collisionless stellar disc.

Theoretical models of JH are two dimensional, and have a finite
mean rotation at the disc centre due to a strong positive gradient
of azimuthal velocity dispersion. Stability properties of the disc
strongly depend on the inner boundary conditions of the disc.
We plan to explore new theoretical models that have more
realistic equilibrium properties.

\section*{Acknowledgments}

We thank Rainer Spurzem, Peter Berczik and Toshio Tsuchyia for
their contributions to the project, for their help in running the
models and for lots of hints and clarifying discussions. We thank
the Deutsche Forschungsgemeinschaft for supporting this project by
DFG 436 RUS 17/10/03. AVK also acknowledges the RFBR (07-02-01204).

\end{document}